\documentclass[openacc]{rstransa}%%%%where rstrans is the template name

\titlehead{Research}

\begin{document}

%%%% Article title to be placed here
\title{Balance Laws as Test of Gravitational Waveforms}

\author{%%%% Author details
Lavinia Heisenberg$^{1}$}

%%%%%%%%% Insert author address here
\address{$^{1}$Institute for Theoretical Physics, Philosophenweg 16, 69120 Heidelberg, Germany}

%%%% Subject entries to be placed here %%%%
\subject{Gravitational Wave Physics, Cosmology, Particle Physics}

%%%% Keyword entries to be placed here %%%%

%%%% Insert corresponding author and its email address}
\corres{Lavinia Heisenberg\\
\email{heisenberg@thphys.uni-heidelberg.de}}

%%%% Abstract text to be placed here %%%%%%%%%%%%
\begin{abstract}
Gravitational waveforms play a crucial role in comparing observed signals to theoretical predictions. However, obtaining accurate analytical waveforms directly from general relativity remains challenging. Existing methods involve a complex blend of post-Newtonian theory, effective-one-body formalism, numerical relativity, and interpolation, introducing systematic errors. As gravitational wave astronomy advances with new detectors, these errors gain significance, particularly when testing general relativity in the non-linear regime. A recent development proposes a novel approach to address this issue. By deriving precise constraints - or balance laws - directly from full non-linear GR, this method offers a means to evaluate waveform quality, detect template weaknesses, and ensure internal consistency. Before delving into the intricacies of balance laws in full non-linear general relativity, we illustrate the concept using a detailed mechanical analogy. We'll examine a dissipative mechanical system as an example, demonstrating how mechanical balance laws can gauge the accuracy of approximate solutions in capturing the complete physical scenario. While mechanical balance laws are straightforward, deriving balance laws in electromagnetism and general relativity demands a rigorous foundation rooted in mathematically precise concepts of radiation. Following the analogy with electromagnetism, we derive balance laws in general relativity. As a proof of concept, we employ an analytical approximate waveform model, showcasing how these balance laws serve as a litmus test for the model's validity.
\end{abstract}
%%%%%%%%%%%%%%%%%%%%%%%%%%%

%%%%%%%%%% Insert the texts which can accomdate on firstpage in the tag "fmtext" %%%%%

\begin{fmtext}

%
%\subsection{Insert B head here}
%%%%% Insert B head here
%Subsection text here.
%
%\subsubsection{Insert C head here}
%%%%% Insert C head here
%Subsubsection text here.
%
%\section{Equations}
%
%Sample equations.
%
%%%% Numbered equation
%\begin{align}\label{1.1}
%\begin{split}
%\frac{\partial u(t,x)}{\partial t} &= Au(t,x) \left(1-\frac{u(t,x)}{K}\right)-B\frac{u(t-\tau,x) w(t,x)}{1+Eu(t-\tau,x)},\\
%\frac{\partial w(t,x)}{\partial t} &=\delta \frac{\partial^2w(t,x)}{\partial x^2}-Cw(t,x)+D\frac{u(t-\tau,x)w(t,x)}{1+Eu(t-\tau,x)},
%\end{split}
%\end{align}
%
%\begin{align}\label{1.2}
%\begin{split}
%\frac{dU}{dt} &=\alpha U(t)(\gamma -U(t))-\frac{U(t-\tau)W(t)}{1+U(t-\tau)},\\
%\frac{dW}{dt} &=-W(t)+\beta\frac{U(t-\tau)W(t)}{1+U(t-\tau)}.
%\end{split}
%\end{align}
%
%%%%% Unnumbered equation
%\begin{eqnarray}
%\frac{\partial(F_1,F_2)}{\partial(c,\omega)}_{(c_0,\omega_0)} = \left|
%\begin{array}{ll}
%\frac{\partial F_1}{\partial c} &\frac{\partial F_1}{\partial \omega} \\\noalign{\vskip3pt}
%\frac{\partial F_2}{\partial c}&\frac{\partial F_2}{\partial \omega}
%\end{array}\right|_{(c_0,\omega_0)}\notag\\
%=-4c_0q\omega_0 -4c_0\omega_0p^2 =-4c_0\omega_0(q+p^2)>0.
%\end{eqnarray}
\end{fmtext}

%%%%%%%%%%%%%%% End of first page %%%%%%%%%%%%%%%%%%%%%

\maketitle

\section{Introduction}\label{sec:Introduction}

The monumental discovery of direct gravitational wave (GW) measurements \cite{LIGOScientific:2016aoc} has ushered in a groundbreaking opportunity to delve into the core of gravitational physics. This achievement, accompanied by subsequent observations, raises the prospect of unveiling uncharted domains that extend beyond the well-established bedrock of general relativity (GR). Envisaging these unexplored territories as a departure from GR, new interactions and degrees of freedom promise to reveal novel phenomena. This advancement has the potential to profoundly enrich our understanding of gravitational dynamics and expand our cosmic comprehension \cite{Heisenberg:2018vsk}.

Central to the analysis of GW observations and their theoretical predictions is the significance of GW waveforms. However, deriving analytical waveforms directly from GR remains an intricate challenge. Instead, these waveforms emerge from a complex amalgamation of post-Newtonian theory, the effective-one-body approach, numerical relativity, and interpolation techniques \cite{Blanchet:2013haa,Buonanno:1998gg,Baumgarte:1998te}. Yet, each of these methods introduces its distinct systematic errors, complicating the synthesis process and potentially introducing inaccuracies. As advanced GW detectors prepare to come online and GW astronomy advances rapidly, the impact of these errors on research becomes undeniable. Ensuring waveform accuracy is paramount in this evolving landscape, particularly when GW observations are employed to scrutinize GR's validity in nonlinear regime. The risk of masking deviations from GR or falsely identifying them due to waveform intricacies underscores the need for an independent mechanism to verify waveform consistency.

A pioneering avenue in this pursuit emerges in recent work \cite{Ashtekar:2019viz}. This innovative approach, articulated in meticulous detail, establishes a foundation rooted in balance laws derived directly from complete, non-linear GR. These precise principles not only offer a standard for evaluating waveform authenticity but also illuminate vulnerabilities inherent in existing waveform templates. Our most recent accomplishment is encapsulated in the comprehensive exposition "Gravitational Waves in Full, Non-linear General Relativity" \cite{DAmbrosio:2022clk}, which serves as the cornerstone for our ongoing efforts.
The potency of these balance laws and their utility in evaluating waveform models are key focal points. This work aims to elucidate these facets as its primary objective.

The article is structured as follows: Section~\ref{sec:GravitationalWaveforms} encapsulates an overview of the factors contributing to errors and limitations in waveform models within the realm of general relativity. Section~\ref{sec:BalanceLawsClassicalMechanics} initiates with an elaborate mechanical analogy. We employ a dissipative mechanical system, specifically a ball's motion down a frictional half-pipe, governed by a non-linear equation. This example showcases how mechanical balance laws serve as a yardstick for evaluating the accuracy of approximate solutions in representing the complete physical scenario. Utilizing a small-angle approximation, we derive an approximate analytical solution and subsequently evaluate its faithfulness in capturing the underlying physics through the unapproximated mechanical balance laws. In Section ~\ref{sec:BalanceLawsGR}, we'll explore how the derivation of balance laws in electromagnetism and general relativity differs from the straightforward mechanical formulation, demanding a rigorous foundation based on mathematical concepts of radiation. Starting with an electromagnetism analogy, we'll introduce key ideas, including conformal completion, the Newman–Penrose formalism, Carter–Penrose diagrams, and the profound Peeling Theorem. After deriving the complete set of balance laws in general relativity, we'll apply them as a litmus test to evaluate the accuracy and validity of a specific approximate waveform model, serving as a proof of concept. Lastly, in Section~\ref{sec:Conclusion}, we will provide a conclusion and offer insights into future directions for this research. This work centers on the examination of balance laws within the domain of general relativity and their utility in scrutinizing gravitational waveform models. It's worth noting that analogous balance laws can also be derived in broader theories of gravity that extend beyond general relativity.

%--------------------------------------------------------------------
%	Gravitational Waveforms
%--------------------------------------------------------------------
\section{Gravitational Waveforms}\label{sec:GravitationalWaveforms}
At the forefront of gravitational waves stands the central figure of black holes. These enigmatic entities constitute regions of spacetime where gravity's grip is so potent that not even light can break free. This concept is a direct outcome of General Relativity's principles. The Schwarzschild black hole, a vacuum solution to Einstein's field equations, embodies an isolated, spherically symmetric, and static spacetime configuration. A spacetime is deemed spherically symmetric when the metric's Lie derivatives with respect to rotations yield a vanishing result
\begin{eqnarray}\label{sphericalspacetime}
\mathcal{L}_{\mathcal{R}_1}g_{\mu\nu} & = 0 \\
  \mathcal{L}_{\mathcal{R}_2}g_{\mu\nu} & = 0 \\
  \mathcal{L}_{\mathcal{R}_3}g_{\mu\nu} & = 0\;,
\end{eqnarray}
where $\mathcal{L}$ stands here for the Lie derivative and $\mathcal{R}_i$ represent the generators of rotations
\begin{eqnarray}\label{rotationsgenerator}
 \mathcal{R}_1 &=& \phantom{-}\sin(\phi) \,\partial_\theta + \frac{\cos(\phi)}{\tan(\theta)}\,\partial_\phi\\
 \mathcal{R}_2 &=& -\cos(\phi)\,\partial_\theta + \frac{\sin(\phi)}{\tan(\theta)}\, \partial_\phi\\
 \mathcal{R}_3 &=& -\partial_\phi\;.
\end{eqnarray}
In the realm of spacetime, the term "stationary" designates a scenario where a timelike vector field $\xi_\text{T}$ can be identified, fulfilling the condition $\mathcal{L}_{\xi_\text{T}}g_{\mu\nu} = 0$. This property encapsulates the essence of "time-translation invariance," manifested as the transformation $t\mapsto t + c$. A vector field that is time-like is inherently situated within either the future or past light cone of an event, thereby indicating its orientation in the "time direction". A spacetime is termed static when a timelike vector field $\xi_\text{T}$ not only conforms to $\mathcal{L}_{\xi_\text{T}}g_{\mu\nu} = 0$ but also stands orthogonal to a spacelike hypersurface. Such a static spacetime embodies the concept of "invariance under time inversion" $t\mapsto -t$. Importantly, static spacetimes encompass the notion of stationarity, though the reverse is not necessarily true. Static solutions pertain to objects devoid of rotation, while stationary solutions characterize entities with rotational attributes. Upon imposing both stationarity and rotational symmetry upon the metric, a notable restriction is placed on the permissible form of the metric\footnote{In scenarios where additional degrees of freedom are present, such as an independent connection or additional scalar or vector degrees of freedom, the application of vanishing Lie derivatives to these fields leads to the emergence of constrained forms. See for instance \cite{DAmbrosio:2021zpm,Heisenberg:2017xda,Heisenberg:2018vti} for some specific examples. }
\begin{eqnarray}\label{sphericalstationarymetric}
g_{\mu\nu} =
\begin{pmatrix}
  g_{tt}(r) & g_{tr}(r) & 0 & 0 \\
  g_{tr}(r) & g_{rr}(r) & 0 & 0 \\
  0 & 0 & g_{\theta\theta}(r) & 0 \\
  0 & 0 & 0 & g_{\theta\theta}(r)\, \sin^2(\theta)
\end{pmatrix}\;.
\end{eqnarray}
Subsequently, the next step involves incorporating this restricted metric ansatz into the vacuum Einstein field equations $G_{\mu\nu} = 0$. To derive solutions for the Einstein field equations, it is imperative to establish appropriate boundary conditions. Initially, we demand asymptotic Minkowski behavior for the spacetime $\lim_{r\to +\infty}g_{\mu\nu}(r) = \eta_{\mu\nu}$. Additionally, we enforce the recovery of Newton's law in regions distant from the source $\ddot{x}^{i} = -\left\{i\atop tt\right\} = \frac{G\,M}{r^2}$. Although boundary conditions play a crucial role, they don't fully determine the solution. There remains a degree of gauge freedom, which is important to acknowledge. It's worth highlighting that various gauge selections can yield solutions with distinct appearances. The Schwarzschild gauge and the Painlev\'e-Gullstrand gauge look for instance as
\begin{eqnarray}\label{schwarzschildmetric}
\mathrm{d}s_\text{Sch}^2 &= -\left(1-\frac{2M}{r}\right)\mathrm{d}t^2 + \left(1-\frac{2M}{r}\right)^{-1}\mathrm{d}r^2 + r^2\,\mathrm{d}\Omega^2\\[5pt]
\mathrm{d}s_\text{PG}^2 &= -\left(1-\frac{2M}{r}\right)\mathrm{d}t^2 + 2\sqrt{\frac{2M}{r}} \mathrm{d}t\,\mathrm{d}r + \mathrm{d}r^2 + r^2\,\mathrm{d}\Omega^2\;.
\end{eqnarray}
In the realm of gravitational wave physics, there is a central theorem for spherically symmetric solutions: The Birkhoff theorem. It asserts that any spherically symmetric solution stemming from Einstein's vacuum field equations inevitably converges to the Schwarzschild solution, which is inherently static. A crucial implication of this theorem is that spherically symmetric, pulsating stars do not emit gravitational waves.

Stationary black hole solutions within the framework of GR are classified into four important categories. The first category encompasses the Schwarzschild solutions, where the black hole's properties are solely determined by its mass, denoted as $M$. In the second class, we encounter the Reissner-Nordström solutions, characterized not only by the mass $M$ but also by an associated electric charge $Q$. The third pivotal category revolves around the Kerr solution, portraying a rotating black hole existing in a vacuum. This solution is defined by two essential parameters: the mass $M$ and the angular momentum $J$. Notably, the Kerr solution unveils both an event horizon and an ergosphere, the latter being the region beyond the event horizon where the frame-dragging phenomenon emerges due to the black hole's rotation. The fourth category comprises the Kerr-Newman solutions, an extension of the Kerr solution that incorporates an electric charge $Q$ as well. These solutions eloquently depict the intricacies of a rotating, charged black hole. This leads us to the No-hair theorem. According to this theorem, every stationary black hole solution derived from the Einstein field equations coupled with Standard Model matter can be fully described by a mere trio of externally measurable parameters: Mass $M$, charge $Q$, and angular momentum $J$.

Hence, to generate and detect gravitational waves, the merger of black holes is essential. When two black holes coalesce, the system becomes distinctly non-spherically symmetric and non-stationary, rendering the no-hair theorem inapplicable. This merger process gives rise to the emission of gravitational wave signals. Gravitational waves manifest as distortions in the fabric of spacetime, inducing alterations in both lengths and geometries. Although confronted with the formidable task of measuring exceedingly minuscule variations on the scale of $\Delta L = 10^{-18}$ meters, we have successfully achieved direct measurements of gravitational waves \cite{LIGOScientific:2016aoc}. When grappling with the measurement of these gravitational wave signals, pivotal inquiries emerge: How can we deduce details regarding the masses of the binary system? How do we ascertain the distance and sky location of said binary system? How do we quantify the overall power emission? How can we subject the theory of gravity to rigorous testing? The solutions to these inquiries are encapsulated in the gravitational waveforms. As an example, by analyzing the frequency evolution of the waveform, one can deduce the masses involved. Utilizing both the amplitude and mass information, the distance can be inferred. The time of arrival, combined with amplitude and phase data at the detectors, provides insight into the binary system's location in the sky, among other aspects.

\begin{figure}[hbt!]
	\centering
	\includegraphics[width=0.7\columnwidth]{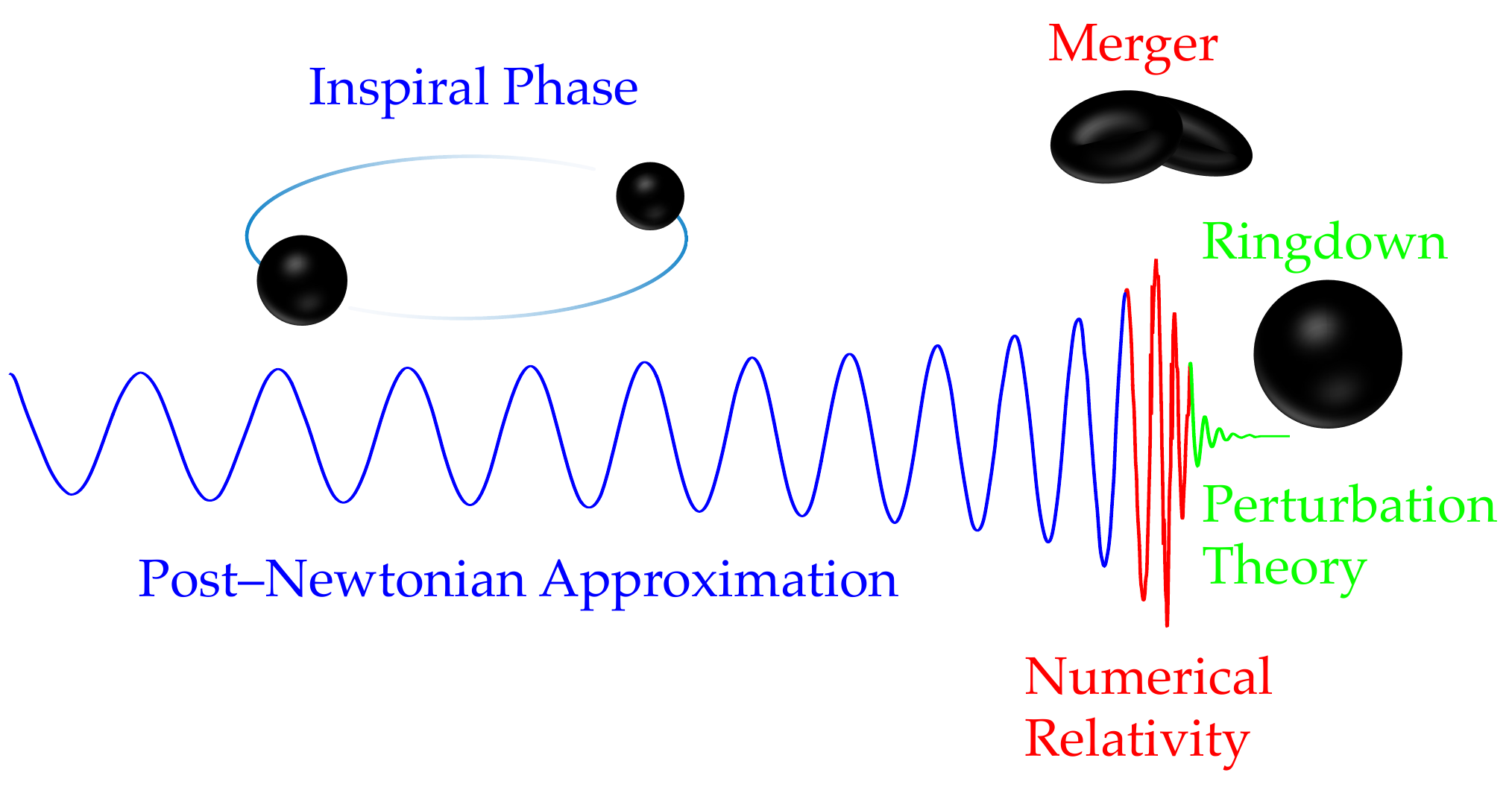}
	\caption{\protect As discussed in the main text, distinct phases of a waveform demand diverse approximation techniques: the Post-Newtonian Approximation for the inspiral phase, Numerical Relativity for the strongly non-linear merger phase, and Perturbation Theory for the subsequent ringdown phase. Achieving a complete waveform entails the amalgamation of these approximated phases into a cohesive whole. This glueing together introduces errors and inaccuracies in the waveform models. \hspace*{\fill}}
	\label{fig:Waveform}
\end{figure}

Solving Einstein's field equations to generate gravitational waveforms is a challenging endeavor due to the complexity of these equations. As a result, describing the coalescence of binary systems involves delineating four distinct sequential phases: the early inspiral, late inspiral, merger, and ring-down. During the early inspiral phase, when the compact objects are still relatively distant, reasonably accurate results can be attained by employing linearized theory. This involves perturbatively expanding the metric as $g_{\mu\nu}=\bar{g}_{\mu\nu}+\delta g_{\mu\nu}$, providing essential information about the propagation speed of gravitational waves and the gravitational constant at the linear order. As the compact objects lose energy in the form of gravitational waves, they draw closer, causing the orbit's frequency to increase. Using the linear theory, estimations for the time of coalescence and the frequency band of emitted gravitational waves can already be derived. As the binary system enters the late inspiral phase, a more precise analytical description becomes necessary. This phase can be tackled through either the Post-Newtonian (PN) expansion \cite{Blanchet:2013haa} or the application of effective field theory techniques \cite{Buonanno:1998gg}. The subsequent ring-down phase also lends itself to a perturbative analytical approach, manifesting as damped oscillations represented by a superposition of quasi-normal modes \cite{Berti:2009kk}. Between these phases lies the merger phase, which defies perturbative analysis and necessitates numerical simulations. To assemble a comprehensive gravitational waveform, these distinct phases must be seamlessly integrated using various computational methods (see Figure \ref{fig:Waveform}). This intricate amalgamation encapsulates the essence of characterizing the complex journey of binary coalescence.

Numerous waveform models are available, encompassing a range of approaches. This non-comprehensive compilation encompasses SXS waveform models, both Pure and Hybridized NR Surrogate models, Effective One Body models, IMPRPhenomD, IMPRPhenomXPHm, IMPRPhenomTPHM, along with a plethora of derived models and more \cite{Boyle:2019kee,Damour:2008qf,Damour:2009kr,Zlochower:2005bj,Pretorius:2005gq,Pan:2013rra,Blackman:2017pcm,Husa:2015iqa,Khan:2015jqa,Boyle:2019kee,Ossokine:2020kjp,Varma:2018mmi}. Of significance to our inquiry is the shared essence and distinctive traits of these models: 
\begin{itemize}
\item Waveform models stand as indispensable instruments for both GW detection and the estimation of parameters for merging binary systems. 
\item These models hinge on a set of ten intrinsic (the 2 masses of the constituents, 6 components representing the spins, 1 parameter for the eccentricity of orbit and 1 parameter for the orientation of orbit) and four extrinsic parameters (luminosity distance from source and 3 parameters determining the orientation). 
\item While each model simplifies certain aspects of physics (for instance Effective One Body models assume a small mass–ratio of the binary and many of the older model neglect the so–called GW memory effect), they uniformly require insights from numerical relativity to capture the merger phase—a computationally costly endeavor. Given the resource-intensive nature of numerical relativity simulations, the coverage of parameter space is inherently limited, allowing for only a finite number of data points. Consequently, each model necessitates the utilization of interpolation techniques to bridge the gaps between these discrete numerical relativity data points.
\end{itemize}

Every approximation technique inherently introduces errors that can propagate through the waveform. Given the pivotal role of waveform models in detecting gravitational wave events and extracting information from the signals, a series of fundamental questions emerges: Among the multitude of models available, which can be relied upon for the analysis of observational data? Which model aligns closest with the predictions of full, non-linear GR regarding compact binary coalescence physics? Can we empirically determine which model yields the smallest error and offers the most accurate approximation? The resounding answer is affirmative. This can be achieved through the utilization of the "Balance Laws."

Gaining an understanding of how gravitational waves are characterized within the realm of full, non-linear theory provides the avenue to construct precise mathematical assessments for waveform models, these balance laws. This concept is advanced in the pioneering study  \cite{Ashtekar:2019viz}. The underlying principle of the balance law approach is straightforward: Binary systems dissipate energy by emitting gravitational waves. The cumulative energy loss must equate to the cumulative energy carried away by these waves. This energy equilibrium offers a means to scrutinize and contrast various models. For a beginner-friendly introduction to the classical scenario of gravitational waves within the framework of full, non-linear general relativity we refer to  \cite{DAmbrosio:2022clk}.

%--------------------------------------------------------------------
%	Balance Laws in Classical Mechanics
%--------------------------------------------------------------------
\section{Balance Laws in Classical Mechanics}\label{sec:BalanceLawsClassicalMechanics}
\begin{figure}[hbt!]
	\centering
	\includegraphics[width=0.6\columnwidth]{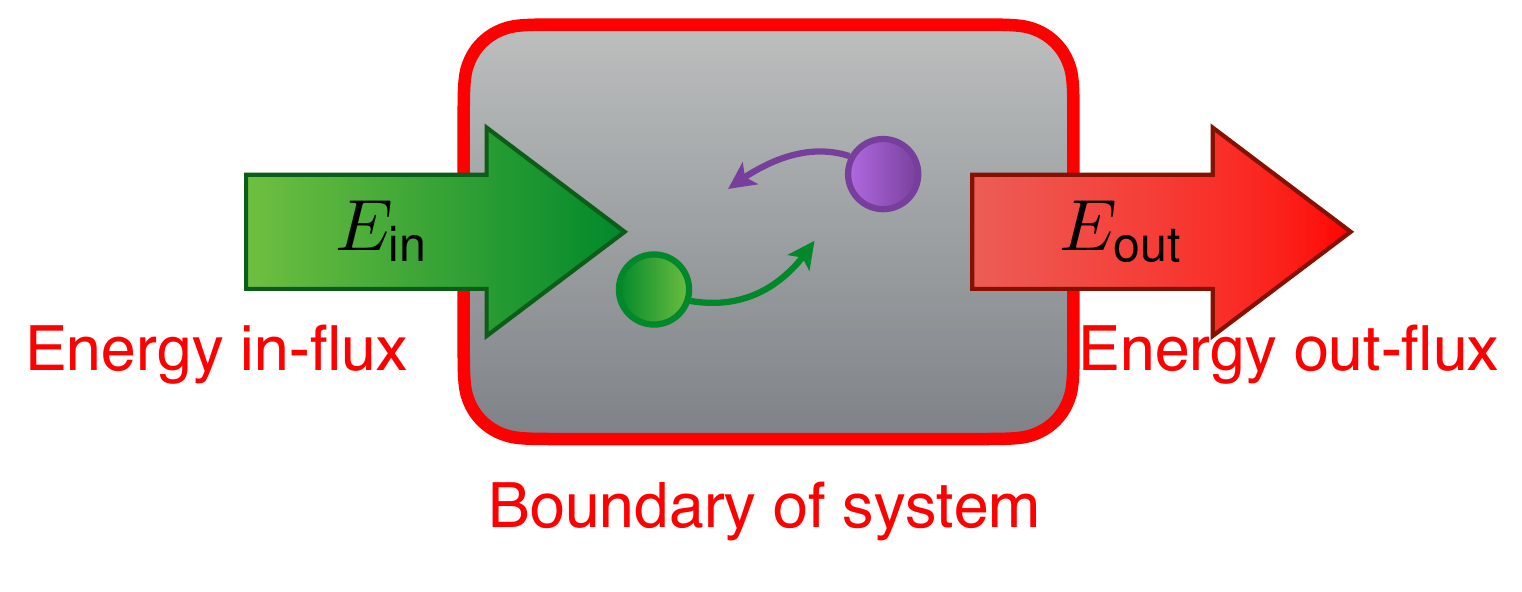}
	\caption{\protect A dissipative mechanical system with a boundary through which an energy in-flux $E_\text{in}$ enters into the system and an energy out-flux $E_\text{out}$ leaves the system. \hspace*{\fill}}
	\label{fig:KlassBL}
\end{figure}

Before delving deeply into the intricacies of the balance laws within the realm of full non-linear general relativity, let's comprehensively illustrate the concept by employing a more detailed mechanical analogy.
Imagine a mechanical system, whether in a relativistic or non-relativistic context, characterized by dissipative dynamics. Now, let's encircle this system with a boundary. In this setup, there will be an energy inflow denoted as $E_\text{in}$ across the boundary, as well as an energy outflow represented by $E_\text{out}$, as depicted in Figure \ref{fig:KlassBL}. It's important to note that the total energy within the system is not conserved due to dissipation. However, a fundamental principle prevails: the cumulative inflow and outflow of energy must precisely balance the overall net gain or loss of energy within the system.

Picture a scenario where we conduct energy measurements on the system at two distinct moments: $t_\text{initial}$ and $t_\text{final}$. These measurements would respectively yield the energy values $E_\text{initial}$ and $E_\text{final}$. From here, a straightforward deduction can be made to establish that
\begin{equation}\label{mechanicalBL}
\color{red}{E_{\text{final}} - E_{\text{initial}} = - \int_{t_\text{initial}}^{t_\text{final}} \frac{\partial L}{\partial t}\,\dd t = E_\text{in} - E_\text{out}}\;.
\end{equation}
This formulation encapsulates the mechanical balance law. On the left side of this equation, we have the total energy change of the system, denoted as $E_{\text{final}} - E_{\text{initial}}$. On the right side, we find the aggregate energy inflow and outflow through the system's boundary $E_\text{in} - E_\text{out}$. In the middle, the equation captures the cumulative inflow and outflow measured through an integral over the time derivative of the Lagrangian $- \int_{t_\text{initial}}^{t_\text{final}} \frac{\partial L}{\partial t}\,\dd t$.

This mathematical principle provides a means to scrutinize approximated solutions to intricate equations of motion. The strategy can be outlined as follows: 
\begin{itemize}
\item Begin by deriving the equations of motion from the Lagrangian $L$. 
\item Introduce simplifying assumptions, approximate equations, or resort to numerical solutions when analytical solutions prove challenging. 
\item Insert the approximated solution into the unaltered, unsimplified balance laws. 
\item The degree to which the balance laws are upheld serves as an indicator of the accuracy with which the approximate solution captures the complete physical scenario.
\end{itemize}

\begin{figure}[hbt!]
	\centering
	\includegraphics[width=0.45\columnwidth]{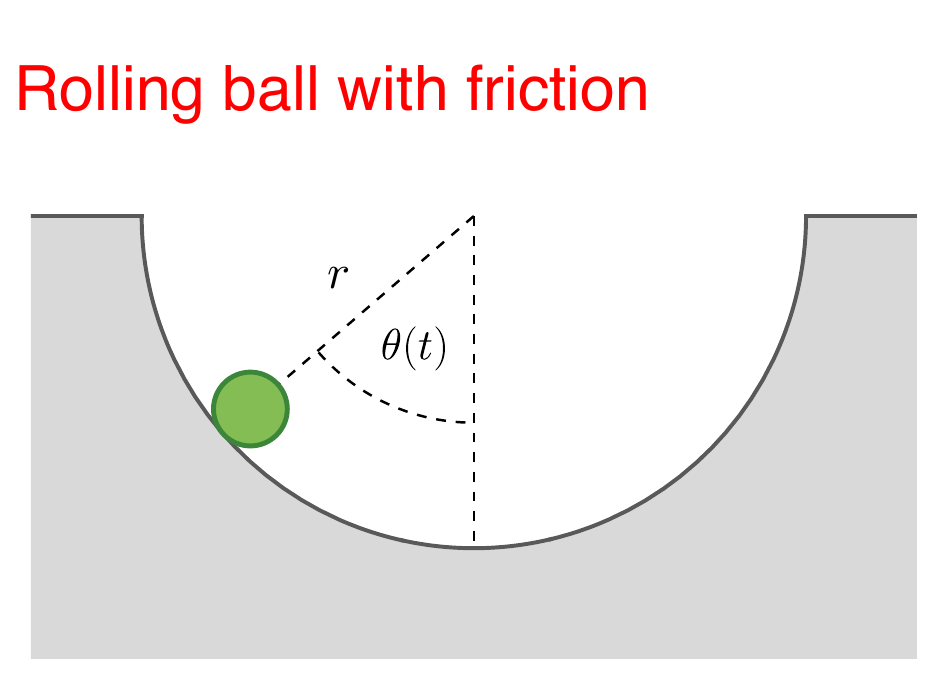}
	\caption{\protect A rolling ball with friction. The ball is described as a point particle and the air resistance is disregarded. The system follows a highly non-linear equations. \hspace*{\fill}}
	\label{RollingBall}
\end{figure}

To exemplify this concept with a tangible instance, let's examine a ball's motion down a frictional half-pipe (see Figure \ref{RollingBall}). For the sake of simplicity, we'll assume the ball to be a point particle and disregard air resistance. Nonetheless, even within this simplified scenario, the equations of motion exhibit considerable non-linearity
\begin{equation}
\ddot{\theta}(t) + g\, \mu\, \dot{\theta}(t) + \frac{g}{r}\, \sin \theta(t) = 0\,,
\end{equation}
with the initial conditions $\dot{\theta}(0) = \omega_0$ and $\theta(0) = \theta_0$. In this equation, the second term signifies the frictional component, while the third term captures the non-linear aspect.

The intricate non-linear equation demands numerical solutions for accuracy. Nonetheless, we can derive an approximate analytical solution through the small-angle approximation. Let us substitute the non-linear term with a linear approximation for small angles
\begin{equation}
\ddot{\theta}(t) + g\, \mu\, \dot{\theta}(t) + \frac{g}{r}\,  \theta(t) = 0\,.
\end{equation}
Now we can solve these approximate equations analytically
\begin{eqnarray}\label{approximatedSol}
\theta(t) &=& \text{e}^{-\frac12 g\, t\, \mu}\left\{\theta_0 \, \cos\left(t\, \sqrt{\frac{g}{r}}\,\sqrt{\left|1-\frac14 g\,\mu^2\,r\right|}\right) \right. \nn\\
&&\left. + \left(\omega_0 + \frac12 g\, \mu\, \theta_0\right) \sqrt{\frac{r}{g}}\frac{1}{\sqrt{\left|1-\frac14 g\,\mu^2\, r\right|}} \sin\left(t\, \sqrt{\frac{g}{r}}\,\sqrt{\left|1-\frac14 g\,\mu^2\,r\right|}\right)\right\} \,.
\end{eqnarray}
The approximate solution seems to describe the physics accurately. But how good is the approximation? By substituting the approximate solution \eqref{approximatedSol} into the unapproximated balance laws \eqref{mechanicalBL} and creating a graphical representation, we can assess the fidelity of the approximated solution in capturing the intrinsic physics. 

\begin{figure}[hbt!]
	\centering
	\includegraphics[width=0.8\columnwidth]{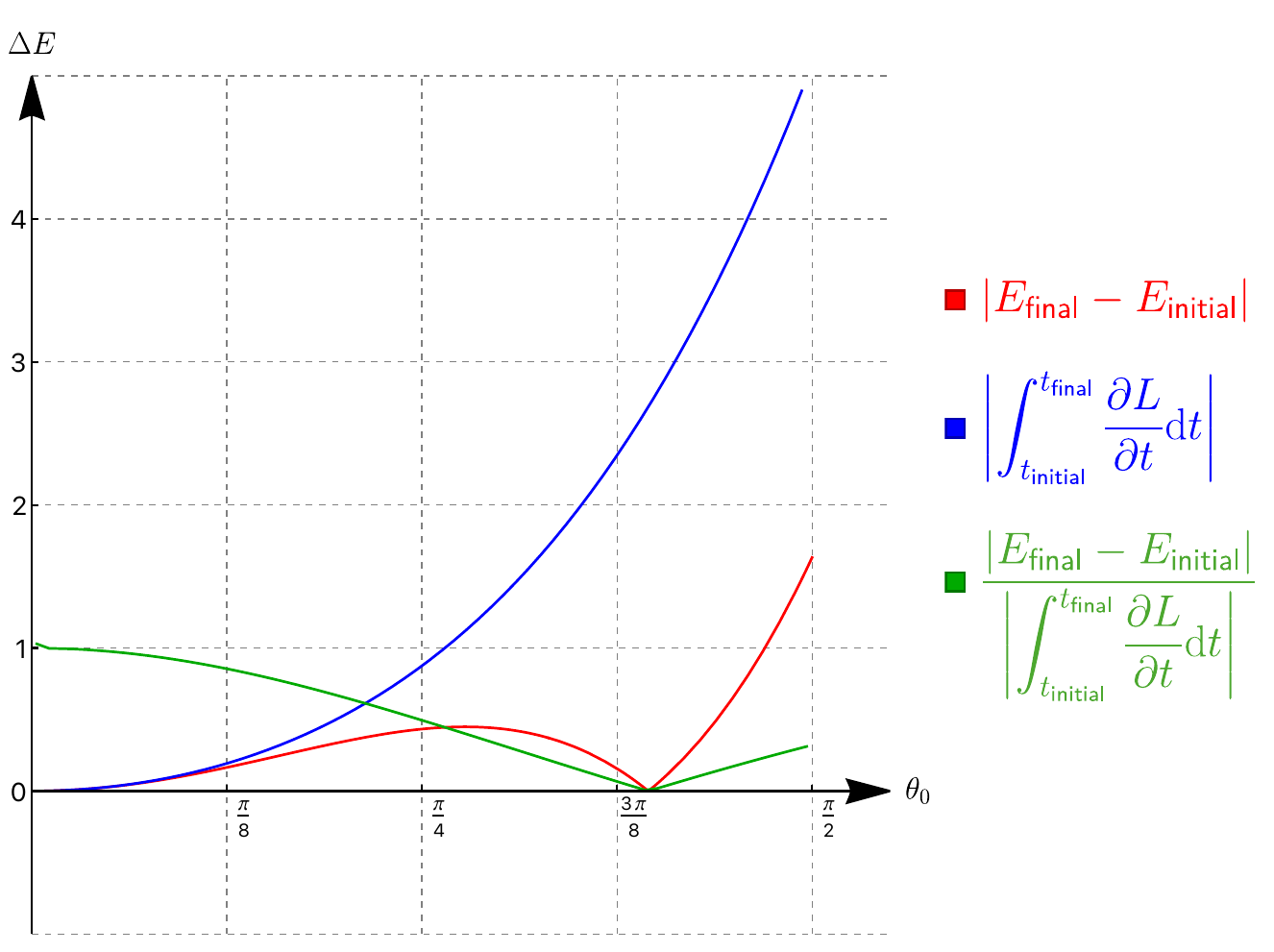}
	\caption{\protect Introducing the approximated solution into the mechanical balance law, we graphically depict the interplay between various components of the balance law. The plot reveals that the balance law aligns with expectations and holds true primarily for small angles. This is a comparison for some fixed parameters $\mu$, $g$, $r$, $\omega_0$ etc. in the approximated solution.  \hspace*{\fill}}
	\label{TestMechBL}
\end{figure}

By systematically exploring constant parameter values, we ascertain the parameter space where the approximated solution effectively captures the fundamental physics. In Figure \ref{TestMechBL}, the analysis demonstrates that the approximation adheres to the balance law for specific parameter values, specifically within the realm of small angles.

We can extend these principles to the realm of general relativity, allowing us to: 1) Compare and enhance waveform models; and 2) Identify the model's superior performance across various regions of parameter space. How this is done is explored in Section \ref{sec:BalanceLawsGR}.

%--------------------------------------------------------------------
%	Electromagnetic Analogy
%--------------------------------------------------------------------
\section{Electromagnetic Analogy}\label{sec:ElectromagneticAnalogy}
Bear in mind that balance laws are precise mathematical relationships inherent to a theory, devoid of approximations. In the pursuit of crafting balance laws for merging binary systems, a comprehensive grasp of gravitational waves within the complete non-linear theory is imperative. For an elaborate and instructive initiation into this topic, we direct readers to \cite{DAmbrosio:2022clk}.

Unlike the straightforward formulation of mechanical balance laws, the derivation of these laws in the realms of electromagnetism and general relativity necessitates a foundation firmly rooted in mathematically rigorous concepts of radiation. Within the framework of the linearized theory, the conceptualization of radiation hinges upon both the wave equation and the intricate dynamics exhibited by its solutions. However, when confronted with a non-linear spacetime of a general nature, the pivotal question arises: How do we discern the presence of radiation within this complex spacetime landscape?

\subsection{The Notion of Radiation}\label{sec:NotionOfRadiation}
\begin{figure}[hbt!]
	\includegraphics[width=0.35\columnwidth]{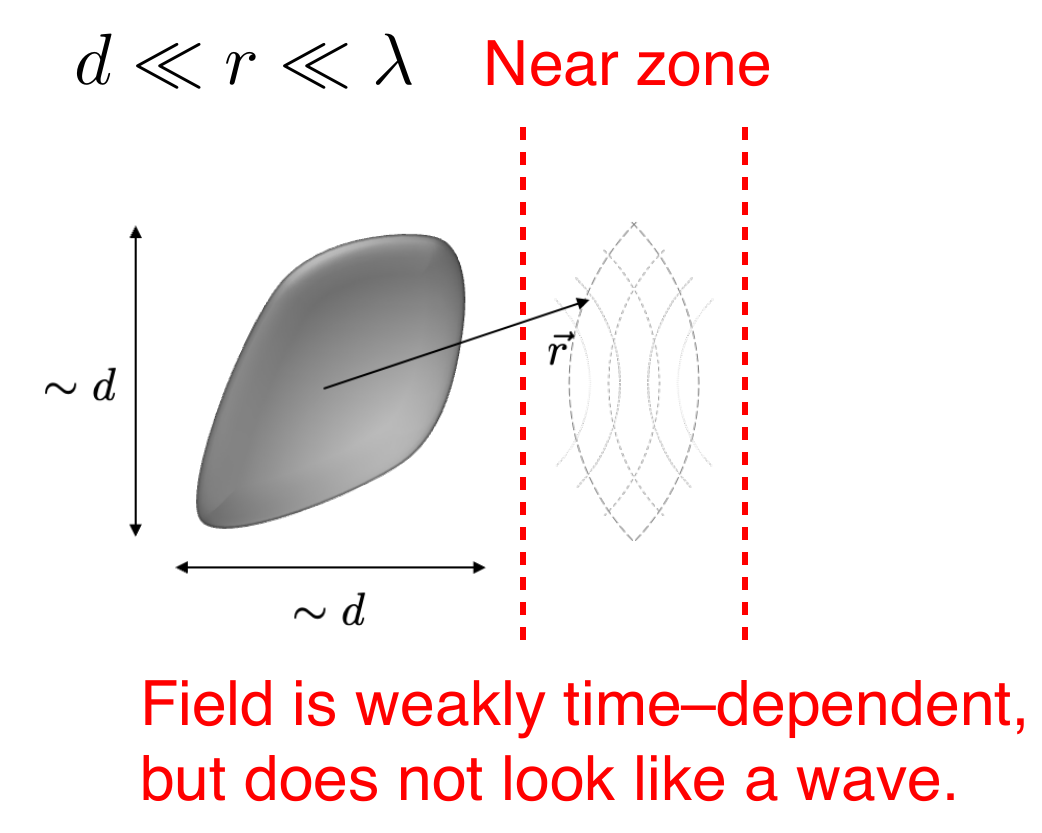}
	\includegraphics[width=0.6\columnwidth]{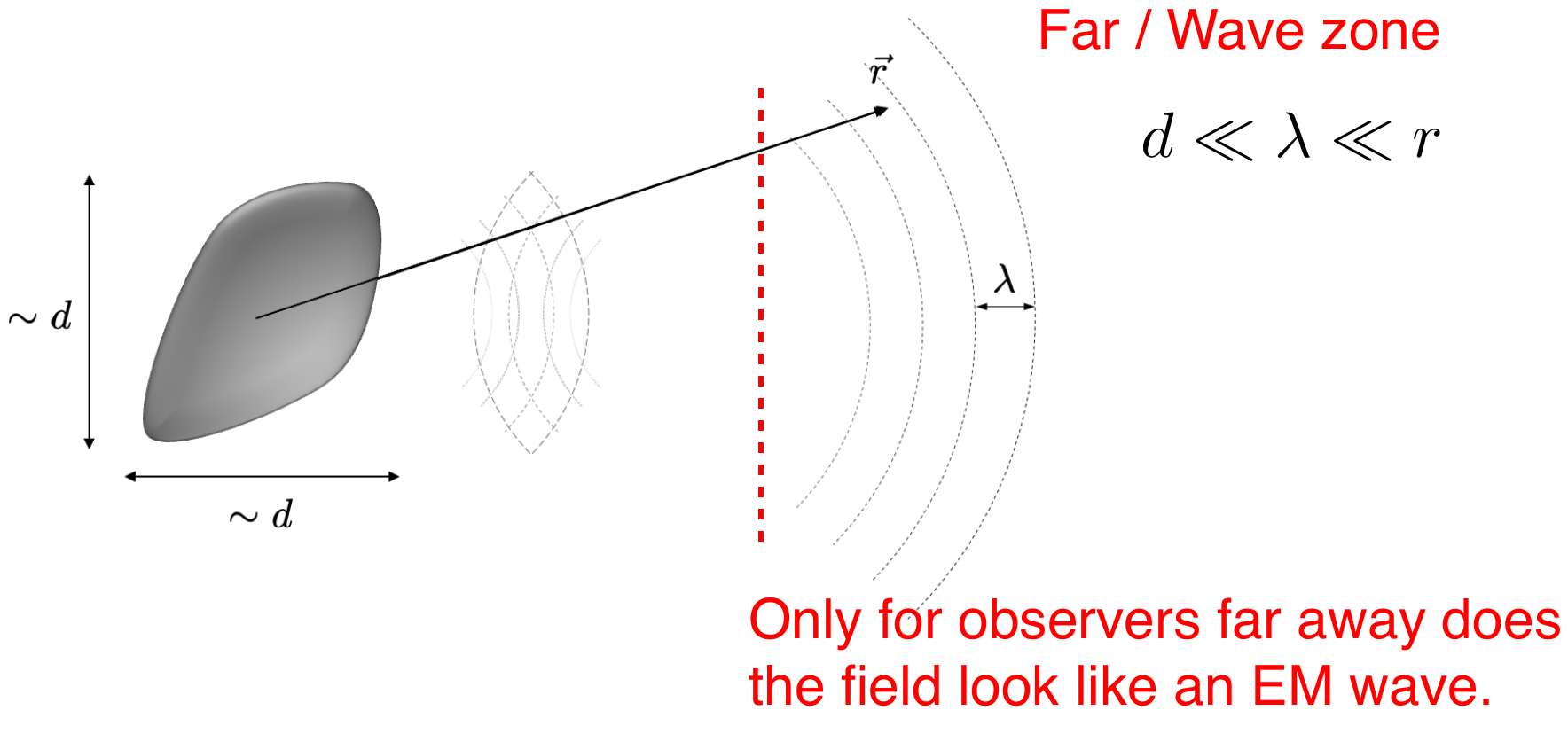}
	\caption{\protect Left:  A pure radiation field can look like a quasi–static field if the observer is located in the near zone. Right: Only for observers far away from the source does the field look like an EM wave.  \hspace*{\fill}}
	\label{NearFarZones}
\end{figure}

Drawing an analogy, let's delve into the scenario within electromagnetism. Deliberating upon a field denoted as $A_\mu$, which elegantly solves the intricate Maxwell's equations, emerges the pressing inquiry: How do we unravel the potential existence of electromagnetic radiation within it? Initially, one might entertain the thought that a cursory inspection of the solution $A_\mu$ to the wave equation would suffice. This line of reasoning, however, stumbles upon immediate roadblocks for two compelling reasons.
Primarily, it's crucial to recall the potency of gauge transformations, capable of engineering ersatz "radiation", an artifice arising when $A_\mu$ manifests oscillatory components in particular gauges, yet these remain entirely bereft of physical significance. Furthermore, the intricacy escalates as we confront the fact that a pure radiation field, under certain circumstances, could adopt the guise of a quasi-static field—particularly when the observer finds their position within the proximity zone. As shown in Figure \ref{NearFarZones}, an essential takeaway to maintain at the forefront of our minds is that the concept of radiation finds its precise delineation solely in the realm of observers positioned a substantial distance from the source. This cardinal principle holds true in the context of gravitational waves as well. 

\begin{figure}[hbt!]
	\centering
	\includegraphics[width=0.8\columnwidth]{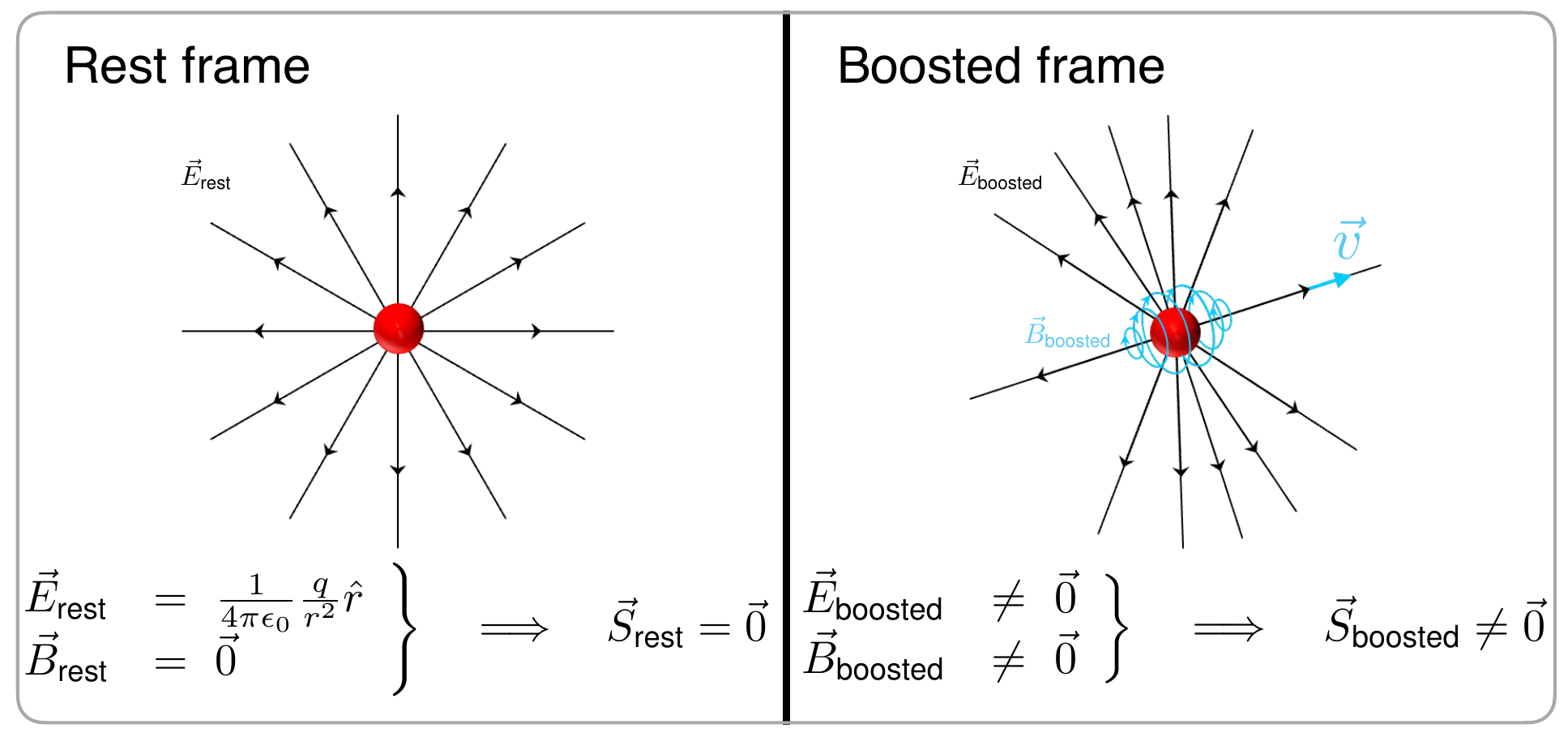}
	\caption{\protect The portrayal of the Coulomb charge from different frames of reference yields the emergence of distinct Poynting vectors. Left is an observer in the rest frame and right is the observer in a boosted frame.  \hspace*{\fill}}
	\label{ChargeFrames}
\end{figure}

A secondary approach towards elucidating the definition of electromagnetic radiation could involve postulating that energy propagation characterizes electromagnetic waves, a phenomenon quantified through the assessment of the Poynting vector $\vec{S} \coloneqq \vec{E}\times\vec{B}$. One might initially conjecture that if $\vec{S}$ is non-zero, it signifies the presence of radiation, whereas if $\vec{S}$ equals zero, it suggests the absence of radiation. it becomes evident that this hypothesis is also flawed. A compelling counterexample arises when examining the behavior of the Coulomb charge as perceived from distinct frames of reference (see Figure \ref{ChargeFrames}). Uniform motion of a charge does not result in radiation. However, our analysis indicates $\vec{S}\neq \vec{0}$ in this scenario. Consequently, we deduce that a mere inspection of the Poynting vector does not suffice in discerning whether a given $A_\mu$ encompasses radiation. Nonetheless, it is imperative to bear in mind the pivotal principle: The precise delineation of radiation holds validity solely in the context of observers positioned at a considerable distance from the source. Certainly, upon the computation of energy flux traversing a 2-sphere $\mathbb{S}^2$ situated at an infinite distance, one arrives for the Coulomb charge at the result
\begin{eqnarray}
&&\lim_{r\to \infty} \oint_{\mathbb{S}^2}\vec{S}_\textsf{rest}\cdot \dd^2\vec{\sigma}\ =\ 0\\
&&\lim_{r\to \infty} \oint_{\mathbb{S}^2}\vec{S}_\textsf{boosted}\cdot \dd^2\vec{\sigma}\ =\ 0\;.
\end{eqnarray}
Henceforth, it becomes evident that the Coulomb charge does not contribute to any energy flux.
It is possible to establish with complete generality that the Poynting vector
\begin{equation}
\vec{S} = \vec{E}\times\vec{B} =  \left(\vec{E}_\text{rad} + \vec{E}_\text{stat}\right) \times \left(\vec{B}_\text{rad} + \vec{B}_\text{stat}\right)
\end{equation}
at far infinity $r\to\infty$ becomes
\begin{equation}
\lim_{r\to\infty}\oint_{\mathbb{S}^2} \left(\vec{E}_\text{rad} + \vec{E}_\text{stat}\right) \times \left(\vec{B}_\text{rad} + \vec{B}_\text{stat}\right)\cdot \dd^2\vec{\sigma} \ = \ \lim_{r\to\infty}\oint_{\mathbb{S}^2} \vec{E}_\text{rad} \times \vec{B}_\text{rad} \cdot \dd^2\vec{\sigma} \neq 0\;.
\end{equation}
Thus, the pivotal takeaway underscores the fact that radiation finds its precise delineation exclusively when the observer is situated at a considerable distance. Moreover, journeying towards "infinity" serves as a mechanism that effectively sieves out all additional components inherent in the electromagnetic field, ultimately presenting us solely with the unadulterated radiative contributions.
Henceforth, the initial query—how to ascertain whether a bestowed vector field $A_\mu$ encompasses radiation—finds its distinct resolution: We must meticulously trace the trajectories of light rays towards the expanse of "infinity" and meticulously scrutinize the electromagnetic field at that remote juncture.

\subsection{Conformally Completed Spacetimes}\label{sec:CCS}
\begin{figure}[hbt!]
	\centering
	\includegraphics[width=0.7\columnwidth]{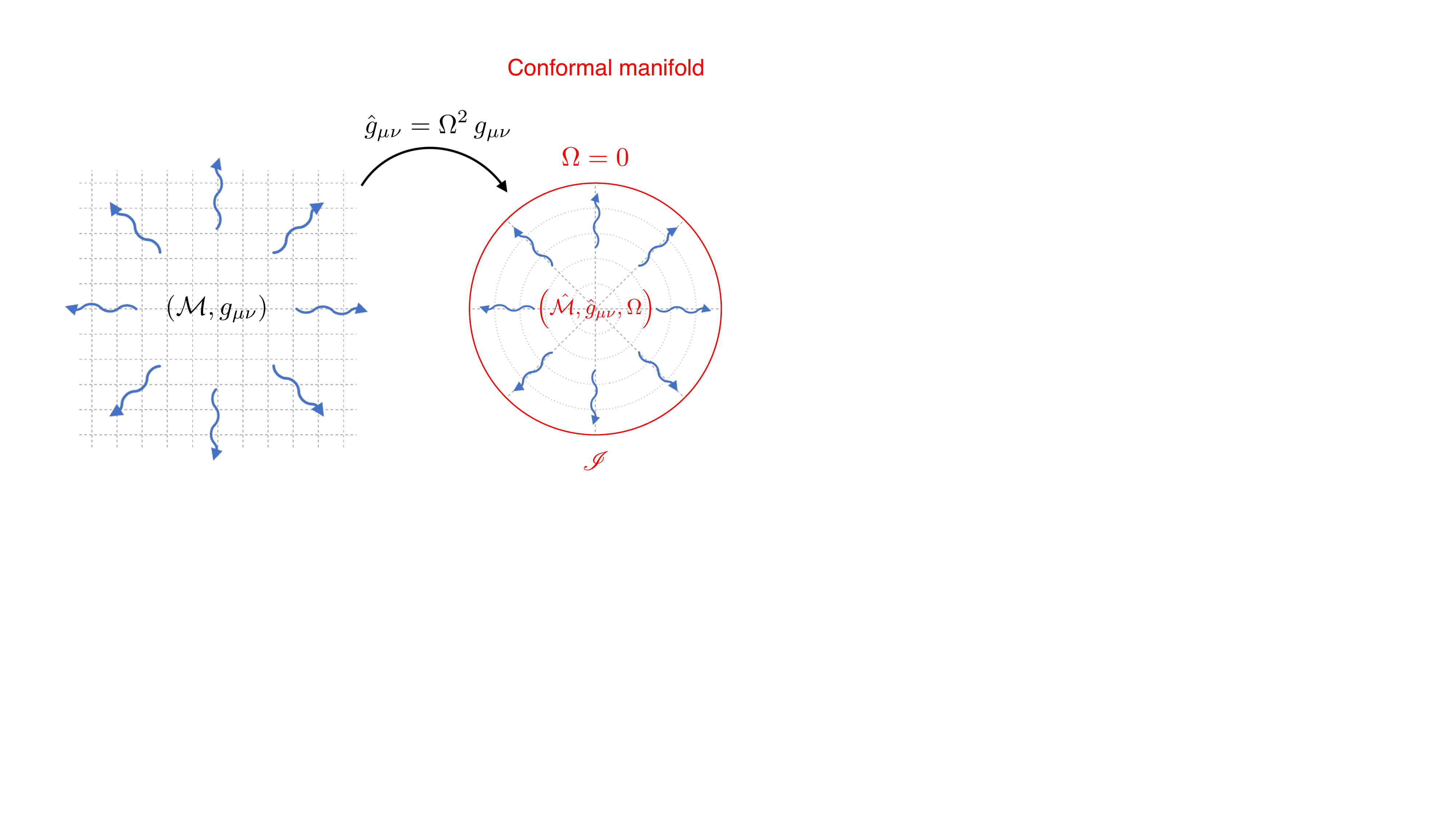}
	\caption{\protect By virtue of the conformal transformation, the notion of infinity is compressed and localized to a finite distance.  \hspace*{\fill}}
	\label{ConformalMap}
\end{figure}

To solidify the concept of tracing light ray trajectories toward infinity, the necessity arises to introduce the notion of achieving spacetime's conformal completion. For a given physical spacetime denoted as $(\mathcal{M}, g_{\mu\nu})$, the concept of a conformally completed spacetime $(\hat{\mathcal{M}}, \hat{g}_{\mu\nu}, \Omega)$ is introduced as $\hat{\mathcal{M}} \coloneqq \mathcal{M}\cup\mathcal{I}$ with $\hat{g}_{\mu\nu} \coloneqq \Omega^2\,g_{\mu\nu}$, where $\hat{\mathcal{M}}$ is the unphysical conformal manifold and $\mathcal{I}$ represents the spacetime boundary. The shift towards the realm of conformal spacetime is effectuated through the utilization of the conformal factor $\Omega$, which assumes a non-zero value across the manifold $\mathcal{M}$ while precisely becoming equal to zero at the boundary $\mathcal{I}$ (see Figure \ref{ConformalMap}).
A conformal transformation exhibits the subsequent attributes: 
\begin{itemize}
\item The inherent causal framework of spacetime remains unaltered. 
\item Notably, light rays continue to traverse null geodesics. 
\item It effectively compresses the concept of infinity to a finite distance.
\end{itemize}

Radiation's departure from the confines of spacetime no longer extends towards the expanse of infinity; rather, it journeys towards the boundary $\mathcal{I}$. This boundary functions akin to a "screen", adeptly capturing and accumulating radiation. It epitomizes the notion of "infinity," ingeniously transformed into a finite proximity. It's crucial to acknowledge that the conformal spacetime, and specifically the boundary/screen structure, are theoretical constructs devoid of physical reality, serving as auxiliary mathematical tools. The advantage bestowed by engaging with the conformally completed spacetime resides in its capacity to bestow precision upon the concept of infinity, permitting the application of differential geometry to the domain $\mathcal{I}$. It's pertinent to underline that $\Omega=0$ can be construed as tantamount to asserting $r=+\infty$ in a certain sense. 

However, the question arises: What precisely does it entail to "pursue light to infinity"? In the context of a Euclidean space ($\mathbb{R}^n, \delta_{\mu\nu}$), there exists a singular category of infinity or asymptotic realm – explicitly, $r\coloneqq \sqrt{\delta_{\mu\nu},x^\mu, x^\nu}\to +\infty$. Within the context of a generic spacetime $(\mathcal{M}, g_{\mu\nu})$, exhibiting the inherent traits of a Lorentzian metric, the discernment of five discrete infinities or asymptotic realms becomes feasible. Each of these distinct regions can be accessed through the pursuit of diverse trajectories or pathways:
i) Timelike: the trajectories traced by material points or observers; ii) Spacelike: indicative of spatial separations at specific instances of time; iii) Null: the pathways charted by massless entities such as light rays.
Timelike trajectories are nestled within the confines of the light cone yet do not coincide with its boundary. Tracing these timelike paths forward enables us to traverse into a domain infinitely distant in the future. Similarly, retracing them into the past guides us to a realm infinitely removed in the past.
Conversely, spacelike pathways extend beyond the light cone, signifying a particular temporal juncture. By traversing these spacelike routes, we embark upon a journey towards a spatial realm of infinite separation.
As for null trajectories, they form the periphery of the light cone. Pursuing these null paths forward leads us to a realm infinitely distanced both in space and time, accessible solely through the passage of light. Similarly, retracing these null paths into the past transports us to an expanse that exists infinitely distant in both spatial and temporal dimensions, again exclusively accessible through the medium of light.

\begin{figure}[hbt!]
	\centering
	\includegraphics[width=1.0\columnwidth]{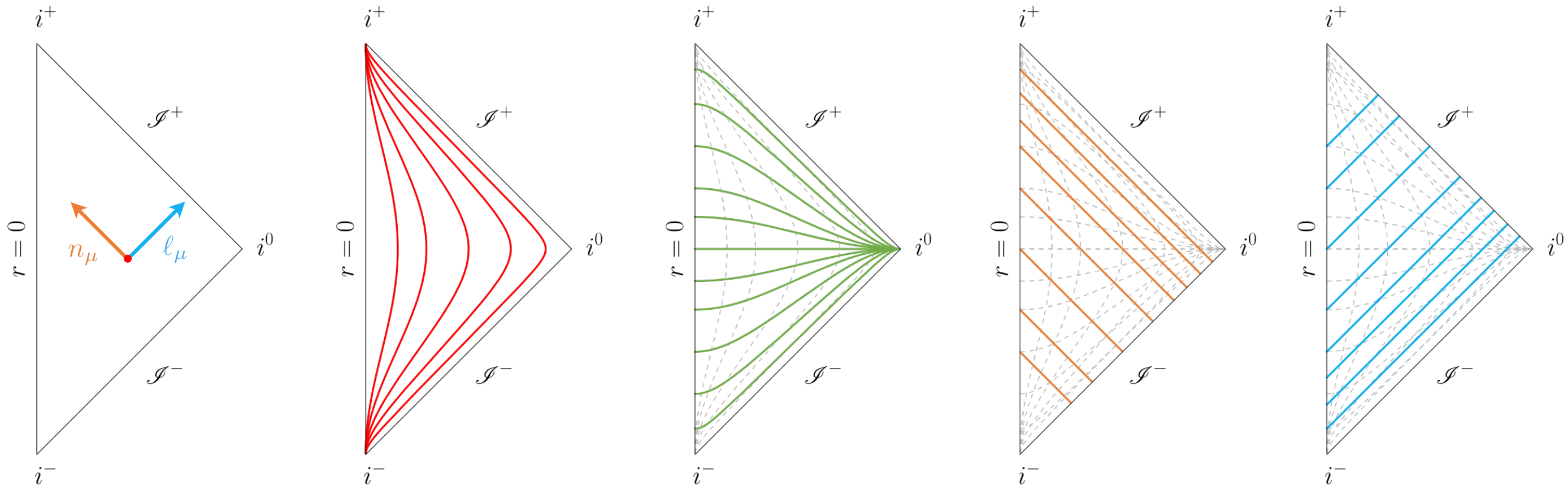}
	\caption{\protect Carter–Penrose diagram of Minkowski space: $i^{-}$ denotes the timelike past infinity: $t\to -\infty$ and $r$ finite. $i^{+}$ stands for the timelike future infinity. $i^0$ is the spacelike infinity: $t$ finite and $r\to +\infty$. $\mathcal{I}^{-}$ represents the past null infinity: $t-r\to -\infty$ such that $t+r$ finite. And $\mathcal{I}^{+}$ is the future null infinity $t+r\to+\infty$ such that $t-r$ finite. $\ell_\mu$ and $n_\mu$ are the tangent null tetrads. The second diagram illustrates trajectories along lines of constant $r$. The third diagram portrays trajectories traced along lines of constant $t$. The forth diagram visualizes the trajectories of ingoing light rays $v \coloneqq t+r =\text{const}$. Finally, the fifth diagram depicts the trajectories of outgoing light rays $u \coloneqq t-r =\text{const}$. \hspace*{\fill}}
	\label{CPDiagram}
\end{figure}

The most apt description for light rays is given in term of $u \coloneqq t - r$ (retarded time) and $v \coloneqq t + r$ (advanced time). It's important to observe that $u=\text{const}$ indicates that $r$ progresses toward the future, while $v=\text{const}$ signifies that $r$ advances toward the past. Within a Lorentzian spacetime, we discern five distinct asymptotic regions:
\begin{itemize}
\item A domain accessible through timelike paths as $t$ approaches $+\infty$.
\item A domain reachable via timelike paths as $t$ approaches $-\infty$.
\item A domain attainable through spacelike paths as $r$ tends towards $+\infty$.
\item A realm accessible through distant-future-traveling light rays.
\item A territory reachable by light rays from the remote past.
\end{itemize}
Regarding the first two regions, for a Minkowski spacetime for instance, we have $\lim_{t\to\pm\infty}\dd s^2 = -\dd t^2 + \dd r^2 + r^2\, \dd \omega^2$. This is bounded, owing to the time-independent nature of the Minkowski metric. In the scenario of the third region, the line element exhibits a divergence due to the explicit $r$ dependence $\lim_{r\to+\infty}\left(-\dd t^2 + \dd r^2 + r^2\, \dd \omega^2\right)$. For the examination of the fourth region, a strategic shift in coordinates from $(t, r, \theta, \phi)$ to $(u, r, \theta, \phi)$ proves to be advantageous. The line element $\dd s^2 = -\dd u^2 -2\,\dd u\, \dd r + r^2\,\dd \omega^2$ diverges for $u=\text{const}$ and $r\to+\infty$. Likewise, the fifth domain $\dd s^2 = -\dd v^2 + 2,\dd v, \dd r + r^2,\dd \omega^2$ where $v=t+r$, displays divergence as $v$ remains constant and $r$ approaches infinity. Evidently, as the limit $r\to+\infty$ is approached, the component $r^2 \dd\omega^2$ within the Minkowski metric showcases divergence. This inherent divergence can be rectified through the application of a conformal transformation $\hat{\eta}_{\mu\nu} = \Omega^2\, \eta_{\mu\nu}$, involving the transition from the original spacetime $(\mathcal{M}, \eta_{\mu\nu})$ to the conformally completed spacetime $(\hat{\mathcal{M}}, \hat{\eta}_{\mu\nu}, \Omega)$. Consequently, by adopting\footnote{The process of conformal completion does not possess inherent uniqueness. An alternative selection could equally have been made $\Omega = \frac{\alpha(\theta, \phi)}{r^n}$, where $n \geq 1$ and $\alpha$ is any smooth, non–zero function.} $\Omega = \frac{1}{r}$, the conformally completed metric $\hat{\eta}_{\mu\nu}$ is rendered finite. Since the conformally completed metric has the line element $\dd \hat{s}^2 = -\frac{1}{r^2}\dd t^2 + \frac{1}{r^2}\dd r^2 + \dd \omega^2$ we have $\lim_{r\to+\infty}\dd \hat{s}^2 =  \dd \omega^2$ finite. Note, that $\Omega = \frac{1}{r}$ implies that $\Omega=0$ indeed corresponds to $r=+\infty$. By omitting the angular dimensions $\theta$ and $\phi$, we can effectively depict Minkowski space within a two-dimensional illustration known as the Carter–Penrose diagram (See Figure \ref{CPDiagram}).

\subsection{Newman–Penrose Formalism}\label{sec:NPFormalism}
For the examination of radiation, it is prudent to employ coordinates $(u, r, \theta, \phi)$, which are better suited for the analysis of outgoing light rays. Furthermore, adopting a cautious approach involves the utilization of null tetrads. These tetrads, specifically tailored to light rays, provide a frame of reference that significantly eases the task of describing radiation phenomena. A frame of reference essentially comprises an ensemble of mutually independent (co-)vectors. Newman and Penrose introduced the subsequent null tetrad for Minkowski space:
\begin{equation}
\ell_\mu \coloneqq -\frac{1}{\sqrt{2}}\left(\delta_\mu{}^t - \delta_\mu{}^r\right)\,, \quad n_\mu \coloneqq -\frac{1}{\sqrt{2}}\left(\delta_\mu{}^t + \delta_\mu{}^r\right)\,,\quad m_\mu \coloneqq \frac{r}{\sqrt{2}}\left(\delta_\mu{}^\theta + i\,\sin\theta\, \delta_\mu{}^\phi\right) \;.
\end{equation}
It is important to note, that $\ell_\mu$ is tangent to $u=\text{const}$ lines and $n_\mu$ tangent to $v=\text{const}$ lines, whereas $m_\mu$ is a complex vector, whose real and imaginary parts are tangent to $2$-sphere. All these vectors are characterized as null vectors, meaning that
\begin{equation}
\eta_{\mu\nu}\, \ell^\mu\, \ell^\nu = 0\,,\quad \eta_{\mu\nu}\, n^\mu\, n^\nu = 0\,,\quad \eta_{\mu\nu}\, m^\mu\, m^\nu = 0 \;.
\end{equation}
The singular scalar products that remain non-vanishing are $\eta_{\mu\nu}\, \ell^\mu\, n^\nu = -1$ and $\eta_{\mu\nu}\, m^\mu\, \bar{m}^\nu = +1$, where $\bar{m}^\nu$ is the complex conjugate of $m^\nu$. The vectors $\{\ell^\mu, n^\mu, m^\mu, \bar{m}^\mu\}$ offer a (complex) basis or reference system that is meticulously tailored to the behavior of light rays. In practical terms, this basis greatly simplifies the task of describing radiation phenomena. For more details see Appendix A of \cite{DAmbrosio:2022clk}.

Before delving into the intricacies of the broader domain of general relativity, it is beneficial to first consider simpler instances like electromagnetic radiation. In this regard, the Newman–Penrose formalism proves valuable. This approach involves expressing the Maxwell 2-form $F_{\mu\nu}$ using the complex basis
\begin{eqnarray}
&&\Phi_2 \coloneqq \phantom{\frac12}F_{\mu\nu}\, n^\mu\,\bar{m}^\nu \\
&&\Phi_1 \coloneqq \frac12 F_{\mu\nu}\, \left(n^\mu\,\ell^\nu + m^\mu\,\bar{m}^\nu\right)\\
&&\Phi_0 \coloneqq \phantom{\frac12}F_{\mu\nu}\, m^\mu\,\ell^\nu\;.
\end{eqnarray}
These are commonly referred to as the Newman–Penrose scalars. Take note that the 2-form $F_{\mu\nu}$ encompasses a total of six components, whereas $\Phi_0$, $\Phi_1$, and $\Phi_2$ constitute complex scalar quantities. Hence, the Newman–Penrose scalars offer a complex portrayal of the six constituent components inherent in $F_{\mu\nu}$. It can be demonstrated that $\Phi_2$ captures the radiative aspects embedded within the electromagnetic field. Meanwhile, the scalar $\Phi_1$ conveys information concerning the Coulombic facets of the electromagnetic field, encompassing aspects such as charge distribution. 

\subsection{The Peeling Theorem}\label{sec:Peeling}
With the implementation of a conformal completion for spacetime, the Newman–Penrose null tetrads undergo a transformation wherein it is mapped to
\begin{equation}
\hat{n}^\mu = n^\mu\,, \qquad \hat{\ell}^\mu = \Omega^{-2}\ell^\mu\,, \qquad \hat{m} = \Omega^{-1}m^\mu\;.
\end{equation}
The observed transformation behavior alone carries the potency to establish a particularly influential theorem: The Peeling Theorem. Our preceding discourse has unveiled several key insights: i) The decay rates of static and radiative components within the electromagnetic field differ; ii) The description of radiation finds its natural home at the juncture of infinity; iii) The establishment of a conformally completed spacetime renders the notion of infinity mathematically rigorous and attainable. These cumulative revelations converge harmoniously within the Peeling Theorem. The representation of the Newman–Penrose scalars can now be articulated both within the context of the physical spacetime and the conformally completed spacetime as follows
\begin{eqnarray}
\Phi_2 &=& \phantom{\frac12}F_{\mu\nu}\, n^\mu\,\bar{m}^\nu \qquad\qquad \qquad \qquad \;\; \hat{\Phi}_2 = \phantom{\frac12}\hat{F}_{\mu\nu}\, \hat{n}^\mu\,\hat{\bar{m}}^\nu\\
\Phi_1 &=& \frac12 F_{\mu\nu}\, \left(n^\mu\,\ell^\nu + m^\mu\,\bar{m}^\nu\right) \qquad \qquad \hat{\Phi}_1 = \frac12 \hat{F}_{\mu\nu}\, \left(\hat{n}^\mu\,\hat{\ell}^\nu + \hat{m}^\mu\,\hat{\bar{m}}^\nu\right) \\
\Phi_0 &=& \phantom{\frac12}F_{\mu\nu}\, m^\mu\,\ell^\nu \qquad \qquad \qquad \qquad \;\; \hat{\Phi}_0 = \phantom{\frac12}\hat{F}_{\mu\nu}\, \hat{m}^\mu\,\hat{\ell}^\nu \;.
\end{eqnarray}
According to the Peeling theorem, as $r$ approaches $+\infty$, the physical Newman–Penrose scalars can be formulated in relation to the conformal scalars in the following manner
\begin{eqnarray}
\Phi_2(u, r, \theta, \phi)\ &=\ \frac{\hat{\Phi}^{\circ}_2(u, \theta, \phi)}{r} + \mathcal{O}\left(\frac{1}{r^2}\right) \\
\Phi_1(u, r, \theta, \phi)\ &=\ \frac{\hat{\Phi}^{\circ}_1(u, \theta, \phi)}{r^2} + \mathcal{O}\left(\frac{1}{r^3}\right)\\
\Phi_0(u, r, \theta, \phi)\ &=\ \frac{\hat{\Phi}^{\circ}_0(u, \theta, \phi)}{r^3} + \mathcal{O}\left(\frac{1}{r^4}\right)\;,
\end{eqnarray}
where $\hat{\Phi}^{\circ}_i(u, \theta, \phi) \coloneqq \left. \hat{\Phi}_i(u, r, \theta, \phi) \right|_{\mathcal{I}^{+}} = \lim_{r\to+\infty}\hat{\Phi}_i(u, r, \theta, \phi)$. It's worth highlighting that $\hat{\Phi}^\circ_2$ serves as a repository for radiation information. Specifically, $\hat{\Phi}^\circ_2=0$ if and only if the absence of electromagnetic radiation is indicated. Conversely, $\hat{\Phi}^\circ_1$ encapsulates the Coulombic modes. In precise terms, $\hat{\Phi}^\circ_2=0$ implies the absence of any net electric and magnetic charges. Furthermore, it's essential to recognize that $\hat{\Phi}^\circ_i$ exhibit gauge invariance, effectively representing the gauge-invariant quantity $F_{\mu\nu}$. These $\hat{\Phi}^\circ_i$ are scalar entities, manifesting a uniform behavior: their vanishing in one coordinate system translates to their vanishing in all coordinate systems.

\begin{figure}[hbt!]
	\centering
	\includegraphics[width=0.7\columnwidth]{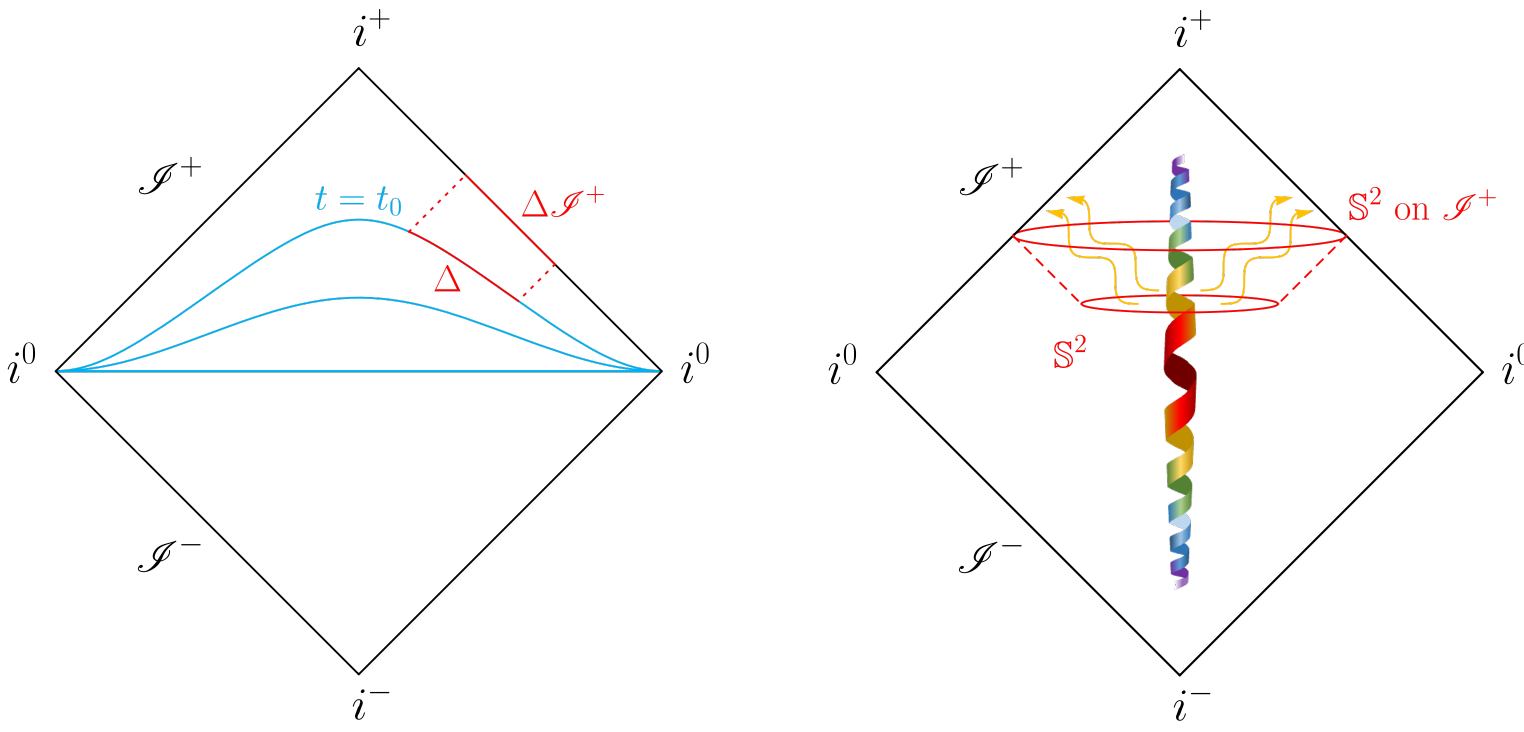}
	\caption{\protect Left: A hypersurface at $t = t_0$ blown-up to infinity. Right: A 2-sphere outside the source blown-up to infinity. \hspace*{\fill}}
	\label{PeelingEM}
\end{figure}

Combining all the elements, we reach the following significant findings:
\begin{itemize}
\item For a given vector field $A_\mu$, the computation of $\hat{\Phi}i$ becomes feasible within the framework of the spacetime completion under conformal transformations $(\hat{M}, \hat{\eta}_{\mu\nu}, \Omega)$.
\item These scalar quantities, $\hat{\Phi}_i$, emerge as robust descriptors of radiation at infinity, transcending both gauge preferences and coordinate choices, thereby ensuring their inherent invariance.
\item The profound Peeling Theorem not only facilitates the calculation of quantities at infinity but also establishes a profound correlation between these quantities and their physical counterparts, thus bridging the gap between mathematical formalism and tangible phenomena.
\end{itemize}
To underscore this final aspect, let us hark back to the notion that the flux of energy and momentum carried by the electromagnetic field through a hypersurface defined at $t = t_0$ can be expressed as follows
\begin{equation}
\int_{t = t_0} T_{\mu\nu}\, n^\mu\, k^\nu\, \dd^3x\;,
\end{equation}
where $n^\mu$ symbolizes the normal vector pointing outwards from the hypersurface, while $k^\nu$ represents the generator responsible for spacetime translations, and the stress-energy tensor is characterized by the expression
\begin{equation}
T_{\mu\nu} \coloneqq F_{\mu\rho}\, F_{\nu \sigma}\,\eta_{\rho\sigma} - \frac14\, \eta_{\mu\nu}\, F_{\rho\sigma} \,F^{\rho\sigma}\;.
\end{equation}
Through blowing-up the hypersurface at $t = t_0$ to extend towards infinity, as visually depicted in Figure \ref{PeelingEM}, we harness the potential of the Peeling Theorem. This theorem, in turn, empowers us to establish a clear link between the flux of energy and momentum through $\mathcal{I}^{+}$, as exemplified by
\begin{eqnarray}
E&=& \int_{\Delta\mathcal{I}^{+}} |\hat{\Phi}^\circ_2|^2\, \sin\theta\, \dd u\, \dd \theta\, \dd\phi \\
P_i&=& \int_{\Delta\mathcal{I}^{+}} \alpha_i\,|\hat{\Phi}^\circ_2|^2\, \sin\theta\, \dd u\, \dd \theta\, \dd\phi \;,
\end{eqnarray}
where $\alpha_i = (\sin\theta\, \cos\phi, \sin\theta\, \sin\phi, \cos\theta)$.
Take note of the fact that the manifestations of energy and momentum are intricately tied solely to $\hat{\Phi}^\circ_2$, the scalar entity that encapsulates the essence of radiation. This observation aligns harmoniously with the established truth that $\mathcal{I}^{+}$ can exclusively be accessed through the trajectories of light rays. Now, let's contemplate a 2-sphere positioned beyond the source, as visually represented in the illustration \ref{PeelingEM}, and envision a process of expanding it towards the conceptual realm of infinity. Subsequently, the Peeling Theorem bestows upon us the following outcome
\begin{eqnarray}
Q_\text{electric} &=& -\frac{1}{2\pi} \oint_{\mathbb{S}^2 \textsf{ on } \mathcal{I}^{+}} \textsf{Re}[\hat{\Phi}^\circ_1]\, \sin\theta\, \dd\theta\, \dd \phi\\
Q_\text{magnetic} &=& -\frac{1}{2\pi} \oint_{\mathbb{S}^2 \textsf{ on } \mathcal{I}^{+}} \textsf{Im}[\hat{\Phi}^\circ_1]\, \sin\theta\, \dd\theta\, \dd \phi\;.
\end{eqnarray}
Notice that the electric and magnetic charges are intricately linked solely to $\hat{\Phi}^\circ_1$, the scalar parameter that embodies the essence of Coulombic contributions.

%--------------------------------------------------------------------
%	Balance Laws in General Relativity
%--------------------------------------------------------------------
\section{Balance Laws in General Relativity}\label{sec:BalanceLawsGR}
Having gained a comprehensive grasp of the principles within the realm of electromagnetism, it is now opportune to shift our focus toward the intricacies of general relativity. For an in-depth exploration, we direct the reader to \cite{DAmbrosio:2022clk}. The application of the Newman–Penrose formalism to electromagnetism facilitated the creation of a description of radiation that remains impervious to gauge transformations and independent of specific coordinate choices, in the distant absence of sources. Building on this achievement, our objective now shifts to extending this formalism to the domain of general relativity, focusing on observers positioned far from any sources. Operating under the assumption of a cosmological constant of zero, and in regions devoid of sources where $T_{\mu\nu}=0$, the Einstein field equations take on the following form
\begin{eqnarray}
R_{\mu\nu} - \frac12 R\, g_{\mu\nu} = 0\;.
\end{eqnarray}
These equations can also be expressed in an equivalent manner as
\begin{eqnarray}
R_{\mu\nu}= 0\;.
\end{eqnarray}
Notice that $R_{\mu\nu}= 0$ implies $R=0$, but not $R^\alpha{}_{\mu\nu\rho} = 0$. Both the Ricci tensor and the Ricci scalar emerge as traces inherent within the Riemann tensor. The vacuum field equations can be comprehended as asserting the nullity of these traces within the Riemann tensor. Yet, there persists the trace-free component of the Riemann tensor, which remains distinct from vanishing:
\begin{equation}
C_{\alpha\mu\nu\rho} \coloneqq R_{\alpha\mu\nu\rho} + g_{\mu[\nu}S_{\rho]\alpha} - g_{\alpha[\nu}S_{\rho]\mu}\;,
\end{equation}
with $S_{\mu\nu} \coloneqq R_{\mu\nu} - \frac16 \, R\, g_{\mu\nu}$. The tensors $C_{\alpha\mu\nu\rho}$ and $S_{\mu\nu}$ are denoted as the Weyl tensor and the Shouten tensor, respectively. Note that in the scenario where $R_{\mu\nu}=0$, it follows that $S_{\mu\nu}=0$, consequently leading to
\begin{equation}
R_{\alpha\mu\nu\rho}=C_{\alpha\mu\nu\rho}\;.
\end{equation}
Our deduction establishes that beyond sources, the entirety of information regarding the gravitational field finds its embodiment within the Weyl tensor. Remarkably, the Weyl tensor undertakes a role analogous to that which $F_{\mu\nu}$ undertook in the domain of electromagnetism. In a manner reminiscent of electromagnetism, we can introduce the Newman–Penrose null tetrads ${\ell^\mu, n^\mu, m^\mu, \bar{m}^\mu}$ that conforms to the following conditions: $g_{\mu\nu}\, \ell^\mu\, n^\nu = -1$ and $g_{\mu\nu}\, m^\mu\, \bar{m}^\nu = +1$. All other scalar products become null in value.
Bear in mind that the Newman–Penrose null tetrads, in essence, furnish us with a complex foundation well-suited for characterizing outgoing light rays. Consequently, it becomes inherently feasible to represent the Weyl tensor within this framework, thereby establishing the foundation for introducing the Newman–Penrose scalars
\begin{eqnarray}
\Psi_4 &&\coloneqq C_{\alpha\mu\nu\rho} \, n^{\alpha}\, \bar{m}^\mu\, n^\nu\, \bar{m}^\rho\\
\Psi_3 &&\coloneqq C_{\alpha\mu\nu\rho} \, \ell^{\alpha}\, n^\mu\, \bar{m}^\nu\, n^\rho\\
\Psi_2 &&\coloneqq C_{\alpha\mu\nu\rho} \, \ell^{\alpha}\, m^\mu\, \bar{m}^\nu\, n^\rho\\
\Psi_1 &&\coloneqq C_{\alpha\mu\nu\rho} \, \ell^{\alpha}\, n^\mu\, \ell^\nu\, m^\rho\\
\Psi_0 &&\coloneqq C_{\alpha\mu\nu\rho} \, \ell^{\alpha}\, m^\mu\, \ell^\nu\, m^\rho \;.
\end{eqnarray}
Recall that in four dimensions the Riemann tensor has twenty independent components. 
Its trace (i.e., the Ricci tensor) has ten independent components. The Weyl tensor (trace–less part of Riemann tensor) has $20-10=10$ independent components. Consequently, there have to be five complex Newman–Penrose scalars ( $=2\times 5$ real components). A critical observation demands our immediate attention at this juncture. Our present context places us within a spacetime characterized by curvature, a departure from the familiar realm of Minkowski space. This fundamental alteration implies that the trajectories of light rays cease to adhere to the paths observed within Minkowski space. Furthermore, it's imperative to recognize that the formulation of the Newman–Penrose null tetrad no longer adheres to the same expressions we encountered in the context of Minkowski space. This underscores the profound impact of curvature on the intricate dynamics of spacetime. In pursuit of the establishment of a well-defined Newman–Penrose null tetrad, our focus pivots towards a particular class of spacetimes known as asymptotically Minkowski spacetimes. In qualitative terms, an asymptotic Minkowski spacetime pertains to a spacetime configuration wherein the equation $\lim_{r\to +\infty} g_{\mu\nu} = \eta_{\mu\nu}$ can be endowed with a rigorously defined mathematical interpretation. A paradigmatic illustration of this concept finds embodiment in the Schwarzschild metric, which, when expressed in Schwarzschild coordinates, assumes the form
\begin{equation}
\dd s^2 = - \left(1-\frac{2M}{r}\right) \dd t^2 + \frac{\dd r^2}{1-\frac{2M}{r}} + r^2\, \dd \omega^2\;.
\end{equation}
Observe that for $r\to+\infty$ we have $1-\frac{2M}{r} \to 1$, and therefore $\dd s^2 \approx -  \dd t^2 + \dd r^2 + r^2\, \dd \omega^2$ for $r\to+\infty$. The $r^2\, \dd \omega^2$ term diverges in the same limit. This is why $\lim_{r\to +\infty} g_{\mu\nu} = \eta_{\mu\nu}$ has no precise mathematical meaning. Nevertheless, circumventing this issue becomes attainable when we operate within the framework of the spacetime that has undergone conformal completion. In order to achieve the process of conformal completion for the Schwarzschild spacetime, a judicious strategy involves the transition of coordinates from the familiar $(t, r, \theta, \phi)$ system to the Kruskal coordinates $(U, V, \theta, \phi)$. These coordinates are purposefully aligned with null geodesics and are explicitly defined as
\begin{eqnarray}
U && \coloneqq 
\begin{cases}
  + \text{e}^{-\frac{t-r}{4M}} \sqrt{\left|1-\frac{r}{2M}\right|} & \text{for } r>2M\\
  - \text{e}^{-\frac{t-r}{4M}} \sqrt{\left|1-\frac{r}{2M}\right|} & \text{for } r<2M
\end{cases}\\
V &&\coloneqq \text{e}^{\frac{t+r}{4M}}\sqrt{\left|1-\frac{r}{2M}\right|} \;.
\end{eqnarray}
Expressed in terms of these coordinates, the Schwarzschild metric takes on the form
\begin{equation}
\dd s^2 = -\frac{32\, M^3}{r}\, \text{e}^{-\frac{r}{2M}}\, \dd U\, \dd V + r^2\, \dd\omega^2\;,
\end{equation}
where $r$ is an implicit function of $U$ and $V$. In order to mitigate the issue of divergent behavior, we opt for the conformal factor as $\Omega = \frac{1}{r}$. The metric, following the conformal rescaling, transforms into 
\begin{equation}
\dd \hat{s}^2 = \Omega^2\, \dd s^2= -\frac{32\, M^3}{r^3}\, \text{e}^{-\frac{r}{2M}}\, \dd U\, \dd V + \, \dd\omega^2\;.
\end{equation}
As a result of this manipulation, we are now able to confidently pursue the $r\to+\infty$ limit, leading us to the conclusion that $\lim_{r\to +\infty} \dd \hat{s}^2 = \dd\omega^2$. Interestingly, this outcome aligns precisely with the limit we previously established for Minkowski space. Henceforth, our focus will be directed towards spacetimes wherein, during the process of conformal completion, the condition $\lim_{r\to+\infty} \hat{g}_{\mu\nu},\dd x^\mu, \dd x^\nu = \dd\omega^2$ prevails. This distinctive category will be referred to as asymptotically Minkowski spacetimes. We can now aptly establish the definition of a Newman–Penrose null tetrad within the vicinity of $\Omega = 0$, specifically in proximity to $r=+\infty$. Furthermore, this framework enables us to establish a connection with the linearized theory. The rationale behind this connection is intuitively evident, as within the vicinity of $\Omega = 0$, an asymptotically Minkowski metric increasingly resembles the Minkowski metric. A comprehensive exposition detailing the systematic construction of asymptotically Minkowski spacetimes and the formulation of Newman–Penrose null tetrads suited for such spacetimes is extensively elaborated in \cite{DAmbrosio:2022clk}.

Shifting our focus to the realm of the Peeling theorem within general relativity, a parallel can be drawn with electromagnetism. Similar to the electromagnetic case, it can be demonstrated that the Newman–Penrose scalars, tailored to suit asymptotically Minkowski spacetimes, adhere to a Peeling Theorem. As the radial coordinate $r$ approaches $+\infty$, the physical Newman–Penrose scalars unfold as expressions derived from the conformal scalars
\begin{eqnarray}
\Psi_4(u, r, \theta, \phi)\ &=& \frac{\hat{\Psi}^{\circ}_4(u, \theta, \phi)}{r} + \mathcal{O}\left(\frac{1}{r^2}\right)\\
\Psi_3(u, r, \theta, \phi)\ &=& \frac{\hat{\Psi}^{\circ}_3(u, \theta, \phi)}{r^2} + \mathcal{O}\left(\frac{1}{r^3}\right)\\
\Psi_2(u, r, \theta, \phi)\ &=& \frac{\hat{\Psi}^{\circ}_2(u, \theta, \phi)}{r^3} + \mathcal{O}\left(\frac{1}{r^4}\right)\\
\Psi_1(u, r, \theta, \phi)\ &=& \frac{\hat{\Psi}^{\circ}_1(u, \theta, \phi)}{r^4} + \mathcal{O}\left(\frac{1}{r^5}\right)\\
\Psi_0(u, r, \theta, \phi)\ &=& \frac{\hat{\Psi}^{\circ}_0(u, \theta, \phi)}{r^5} + \mathcal{O}\left(\frac{1}{r^6}\right)\;,
\end{eqnarray}
where $\hat{\Psi}^{\circ}_i(u, \theta, \phi) \coloneqq \left. \hat{\Psi}_i(u, r, \theta, \phi) \right|_{\mathcal{I}^{+}} = \lim_{r\to+\infty}\hat{\Psi}_i(u, r, \theta, \phi)$. This theorem once again offers us the opportunity to conduct calculations within the unphysical, conformally completed spacetime, subsequently translating these outcomes back into the physical spacetime. Moreover, it's noteworthy that the physical $\Psi_4$ exhibits a decay rate of $1/r$, aligning precisely with the anticipated decay pattern of gravitational waves. Consequently, it stands as a reasonable conjecture that $\Psi_4$ indeed encapsulates the essence of gravitational waves. Let us recollect that as $r$ approaches $+\infty$, asymptotically Minkowski spacetimes gradually resemble Minkowski space more closely. This affords us a straightforward means of establishing correspondence with the linearized theory, denoted by $g_{\mu\nu} = \eta_{\mu\nu} + h_{\mu\nu}$, where gravitational waves are characterized by the polarizations $h_+$ and $h_\times$, intrinsic to the metric perturbations. In fact, it is possible to demonstrate
\begin{equation}
\hat{\Psi}^\circ_4 = -\frac12 \left(\ddot{h}_+ - i\, \ddot{h}_\times\right)(u, \theta, \phi)\;.
\end{equation}
Recall that in electromagnetism we had $\hat{\Phi}^\circ_2 = 0$ if and only if there is no radiation. Conversely, within general relativity, we discover that $\hat{\Psi}^\circ_4 = 0$ holds as a necessary and sufficient condition denoting the absence of radiation. Drawing parallels from electromagnetism, where $\hat{\Phi}^\circ_1$ encapsulated Coulombic contributions, akin to net charge, we contemplate if a counterpart exists in general relativity. Is there a scalar entity that encapsulates the concept of net "gravitational charge"? Indeed, there is. The real component of $\hat{\Psi}^\circ_2$ inherently conveys insights into mass distribution. Summing up, we can construct the following qualitative framework: i) Given a solution $g_{\mu\nu}$ to Einstein's equations that satisfies the asymptotic Minkowski condition, we have the means to discern the presence of radiation by evaluating $\hat{\Psi}^\circ_4$. ii) Utilizing the same $g_{\mu\nu}$, we can extract valuable insights into the cumulative mass by computing $\hat{\Psi}^\circ_2$.

Let's consolidate our progress and accomplishments up to this point:
\begin{itemize}
\item Our primary objective revolves around formulating balance laws that pertain to gravitational waves.
\item In pursuit of this objective, our task encompasses comprehending the means to describe gravitational waves within the framework of the full, non-linear theory.
\item Drawing insights from our exploration of radiation in electromagnetism, we discerned that the unequivocal characterization of radiation pertains exclusively to observers positioned infinitely distant from the source.
\item This realization prompted the introduction of two pivotal concepts: the concept of conformal completions of spacetimes and the utilization of the Newman–Penrose formalism.
\item Conformal completions play a crucial role in rendering a well-defined meaning to the concept of "infinity."
\item The Newman–Penrose formalism offers us a method to "trace the paths of light rays" to infinity, thereby furnishing a convenient framework for describing radiation.
\item When presented with a field, whether it's $A_\mu$ or $g_{\mu\nu}$, the ability to ascertain the presence or absence of radiation becomes attainable through the computation of specialized scalars at infinity.
\end{itemize}

\subsection{Exact Balance Laws}\label{sec:BalanceLaws}
Certainly, it's important to remember that balance laws are precise mathematical relationships that hold true within systems where energy is not conserved. In the context of electromagnetism, we observed how to delineate the energy and momentum of the electromagnetic field within the framework of the Newman–Penrose formalism. However, a comparable methodology for accomplishing this in general relativity is yet to be explored. By invoking a generalized interpretation of Noether's theorem, we can develop apt conceptualizations of energy and momentum tailored to the gravitational fields. Let's examine an event that triggers the emergence of gravitational waves. Within the framework of the conformally completed spacetime, this gravitational wave event propagates outward towards $\mathcal{I}^{+}$. The comprehensive flux of energy carried by gravitational waves through the domain of $\mathcal{I}^{+}$ is meticulously defined as
\begin{equation}\label{EnergyGWs}
\mathcal{F}_{0} \coloneqq \frac{1}{4\pi G} \int_{\mathcal{I}^{+}} \dd u\, \dd\omega^2\left[\dot{h}^2_+ + \dot{h}^2_\times - \text{Re}\left(\eth^2\left(\dot{h}_+ - i\,\dot{h}_\times\right)\right)\right](u, \theta, \phi)\;,
\end{equation}
where the operator $\eth^2$ acts as 
\begin{eqnarray}
\eth^2 f = \frac14 (\sin\theta)^{-1}\left(\partial_\theta + \frac{i}{\sin\theta}\partial_\phi\right)(\sin\theta)^{-1}
 \left(\partial_\theta + \frac{i}{\sin\theta}\partial_\phi\right) (\sin\theta)^{-2} f \;.
\end{eqnarray}
We acknowledge that the first two terms in \eqref{EnergyGWs}, i.e. $\dot{h}^2_+ + \dot{h}^2_\times$, embodies the energy content attributed to gravitational waves, which is also encountered in the linearized theory.
The last term $\text{Re}\left(\eth^2\left(\dot{h}_+ - i\,\dot{h}_\times\right)\right)$ is the gravitational wave memory, which we will elaborate on later. Analogously, the momentum flux carried by gravitational waves through 
$\mathcal{I}^{+}$ is characterized by the following definition
\begin{equation}\label{MomentumGWs}
\mathcal{F}_{i} \coloneqq \frac{1}{4\pi G} \int_{\mathcal{I}^{+}} \dd u\, \dd\omega^2\, \alpha_i \left[\dot{h}^2_+ + \dot{h}^2_\times - \text{Re}\left(\eth^2\left(\dot{h}_+ - i\,\dot{h}_\times\right)\right)\right](u, \theta, \phi)\;,
\end{equation}
with $\alpha_i = (\sin\theta\, \cos\phi, \sin\theta\, \sin\phi, \cos\theta)$.
\begin{figure}[hbt!]
	\centering
	\includegraphics[width=1.0\columnwidth]{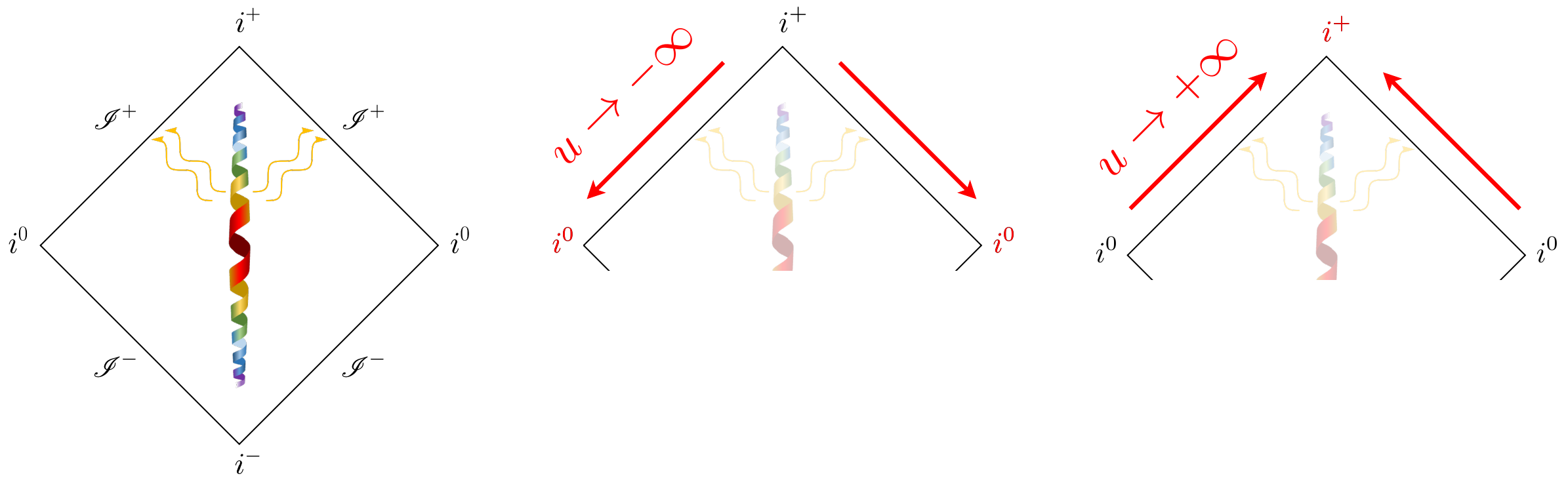}
	\caption{\protect An event gives rise to the emission of gravitational waves, which then propagate away from the source, extending to the far reaches of infinity and eventually reaching the boundary $\mathcal{I}^{+}$. The fluxes \eqref{EnergyGWs} and \eqref{MomentumGWs} reach $\mathcal{I}^{+}$.  \hspace*{\fill}}
	\label{GWsEmission}
\end{figure}

It can be demonstrated that these fluxes can be expressed as integrals of total derivatives with respect to the retarded time $u$. Hence, by employing Gauss' Theorem, the integrals $\mathcal{F}_{A}$ can be reformulated as integrals over the boundaries corresponding to $u = -\infty$ and $u = +\infty$
\begin{equation}\label{BoundaryIntegrals}
\mathcal{F}_A = \lim_{u_0 \to +\infty} \left.\mathcal{P}_A\right|_{u=u_0} - \lim_{u_0\to-\infty} \left.\mathcal{P}_A\right|_{u=u_0}\;
\end{equation}
where we defined 
\begin{equation}\label{BoundaryIntegrals2}
\left.\mathcal{P}_A\right|_{u=u_0} \coloneqq \frac{1}{4\pi G}\oint\dd^2\omega\, A(\theta, \phi)\, \text{Re}\left[\hat{\Psi}^\circ_2 + h_+ \, \dot{h}_+ + h_\times\, \dot{h}_\times\right]\;.
\end{equation}
Equation \eqref{BoundaryIntegrals} provides us with the foundation to articulate the balance laws. Recall that the first term in \eqref{BoundaryIntegrals2}, i.e. $\hat{\Psi}^\circ_2$, encodes information about the masses. Demonstrably, it can be shown that
\begin{itemize}
\item $\lim_{u\to-\infty} \hat{\Psi}^\circ_2 = -G\, M_{i^0}$, where $M_{i^0}$ is the total mass measured on spacelike infinity $i^0$
\item $\lim_{u\to+\infty} \hat{\Psi}^\circ_2 = -G\, M_{i^+}$, where $M_{i^+}$ is the total mass measured on future timelike infinity $i^+$
\end{itemize}
It's important to highlight that these outcomes remain valid solely when the observers stationed at $i^0$ and $i^+$ are at rest relative to the system. In order to advance, it becomes essential to impose rational boundary conditions at both $i^0$ and $i^+$. We shall make the assumption that in the remote past, radiation was absent. This assumption can be articulated as $\dot{h}_+ = 0$ and $\dot{h}_\times = 0$ at $i^0$. In a parallel manner, we shall posit that in the distant future, gravitational waves subside at a specific rate. This supposition can be expressed as $\dot{h}_+ \to 0$ and $\dot{h}_\times \to 0$ at $i^+$.
It's worth noting that these boundary conditions entail $h(u, \theta, \phi) = h_{\pm}(\theta, \phi)\, + \, |u|^{-\epsilon}\, h^{(1)}_{\pm}(\theta, \phi)$ for $u \to \pm \infty$, where $h$ stands for either $h_+$ or $h_\times$ and $\epsilon>0$. These boundary conditions possess sufficient strength to guarantee the absence of radiation in the remote past and future, i.e. $\hat{\Psi}^\circ_4 = 0$ for $u \to \pm \infty$. However, it's important to recognize that these conditions are not of such a stringent nature that they would preclude the presence of the memory effect. 
Additionally, we will make the supplementary assumption that the observer stationed at $i^0$ remains at rest in relation to the source. This assumption, coupled with the established boundary conditions, subsequently leads to
\begin{equation}
\lim_{u_0 \to -\infty}\left.\mathcal{P}_{0}\right|_{u=u_0} \coloneqq \frac{1}{4\pi G}\oint\dd^2\omega\, \, \lim_{u_0\to-\infty}\text{Re}\left[\hat{\Psi}^\circ_2\right]\;.
\end{equation}
The integral $\oint\dd^2\omega$ gives $4\pi$ and $\lim_{u_0\to-\infty}\text{Re}\left[\hat{\Psi}^\circ_2\right]=-GM_{i^0}$. Therefore, for the limit $u_0 \to -\infty$ we obtain
\begin{equation}
\lim_{u_0\to-\infty}\left.\mathcal{P}_0\right|_{u=u_0} = - M_{i^0} 
\end{equation}

The scenario becomes more intricate as we approach the limit $u\to+\infty$. This arises from the fact that we must consider the observer stationed at $i^+$ to no longer remain stationary with respect to the source. This intricacy can be succinctly grasped through a fundamental physical explanation: the source emits gravitational waves that inherently bear momentum. This momentum, carried away by the gravitational waves, exerts an equal-magnitude yet opposing-force momentum onto the source itself. The consequential outcome manifests as the source being propelled into motion, as observed from the vantage point of the initially stationary observer.
This phenomenon, commonly referred to as either "recoil" or "kick," assumes prominence. The velocity of the kick emerges as a crucial astrophysical parameter, accessible through deductions drawn from the characteristics of gravitational waveforms. Kick velocities typically span the range of several thousand kilometers per second, imparting sufficient energy to expel black holes from their galactic hosts.

To culminate in the establishment of the balance laws, it becomes imperative to factor in the circumstance that the aftermath of a collision involving condensed astrophysical entities might exhibit motion at a magnitude of approximately $v_\textsf{kick} \sim \mathcal{O}\left(10^3,\text{kilometers per second}\right)$
This process involves a meticulous examination of the transformation of $\hat{\Psi}^\circ_2$ under Lorentz boosts, leading to the subsequent revelation
\begin{equation}
\lim_{u_0 \to +\infty} \hat{\Psi}^\circ_2 = - \frac{G\, M_{i^+}}{\gamma(v_\textsf{kick})^3\, \left(1 - \vec{v}_\textsf{kick}\cdot\hat{x}\right)^3}\;,
\end{equation}
where $\gamma(v) \coloneqq \frac{1}{\sqrt{1-v^2}}$ is the standard Lorentz factor. It then follows that
\begin{equation}
\lim_{u_0\to +\infty} \left.\mathcal{P}_0\right|_{u=u_0} = - \frac{M_{i^+}}{\gamma(v_\textsf{kick})^3\, \left(1-\vec{v}_\textsf{kick}\cdot \hat{x}\right)}\;,
\end{equation}
with $\hat{x} = \left(\sin\theta\, \cos\phi, \sin\theta\, \sin\phi, \cos\theta\right)$. 

\begin{figure}[hbt!]
	\centering
	\includegraphics[width=1.0\columnwidth]{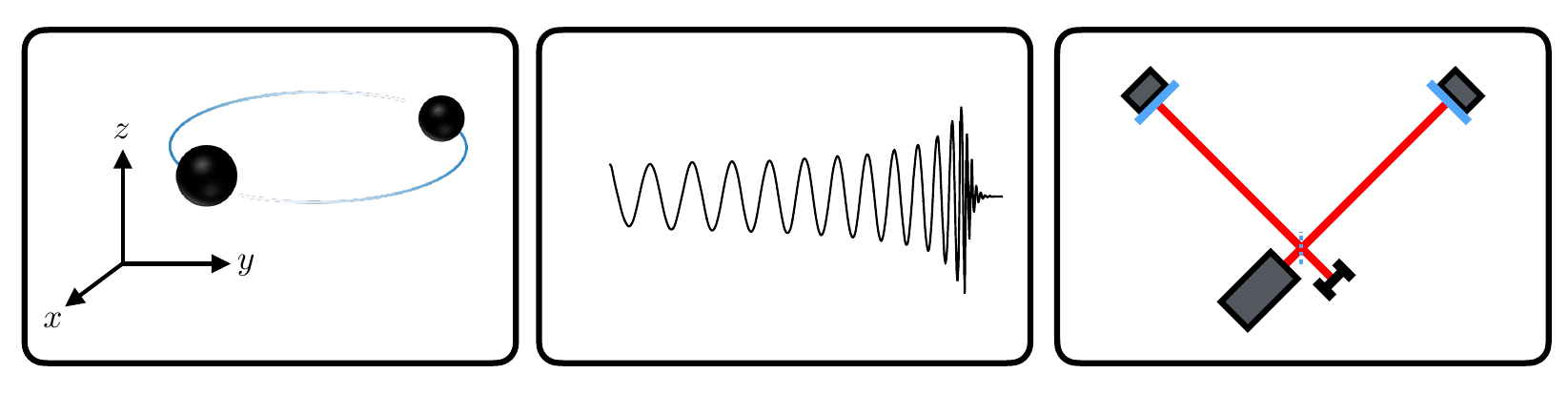}
	\caption{\protect A binary system of compact astrophysical objects emits gravitational waves. In the distant past the observer is at rest relative to the centre of mass of the system. The out–flux of energy is measured by the distant interferometers. \hspace*{\fill}}
	\label{GWsObservation}
\end{figure}

At this juncture, we have reached a pivotal moment where we can definitively articulate the principles governing the balance laws for general relativity.
We consider a binary system of compact astrophysical objects. We choose our coordinate system such that in the distant past, the observer is at rest relative to the centre of mass of the system. Due to the emission of gravitational waves, there is an out–flux of energy.
The discernible signals are captured by interferometers positioned at an effectively "infinite" distance away. These signals, in turn, furnish us with the means to ascertain the following critical parameters:
i) The magnitude of the propulsion velocity, denoted as $\vec{v}_\textsf{kick}$;
ii) The comprehensive mass attribute of the binary entity, symbolized as $M_\textsf{binary} \coloneqq M_{i^0}$;
iii) The magnitude characterizing the remnant's mass, designated as $M_\textsf{remnant} \coloneqq M_{i^+}$.
Recall that the energy flux satisfies the balance law $\mathcal{F}_0 = \lim_{u_0 \to +\infty} \left.\mathcal{P}_0\right|_{u=u_0} - \lim_{u_0\to-\infty} \left.\mathcal{P}_0\right|_{u=u_0}$.
By amalgamating all the components into a coherent whole and reintroducing the relevant physical units, we ultimately arrive at our conclusive result for the balance laws of general relativity
\begin{eqnarray}\label{GRBL}
\color{red}{c^2\left(\frac{M_{\textsf{remnant}}}{\gamma(v_\textsf{kick})^3\,\left(1-\frac{\vec{v}_\textsf{kick}}{c}\cdot\hat{x}\right)^3} - M_{\textsf{binary}}\right) }&\color{red}{=}&   \color{red}{-\frac14\frac{D^2_\textsf{L}\, c^3}{G}\int_{-\infty}^{\infty} \left(\dot{h}^2_+ + \dot{h}^2_\times\right)\mathrm{d} t} \nn\\
&& \color{red}{ + \left.\frac12 \frac{D_\textsf{L}\, c^4}{G}\textsf{Re}\left[\eth^2 \left(h_+ - i\, h_\times\right)\right]\right|_{t=-\infty}^{t=+\infty}}\;.
\end{eqnarray}
Note the striking similarity to the mechanical balance laws in \eqref{mechanicalBL}
\begin{equation}
E_\text{final} - E_\text{initial} = - \int_{t_\text{initla}}^{t_\text{final}} \frac{\partial L}{\partial t}\,\dd t \;.
\end{equation}
In both sets of balance laws, the left-hand side features the energy difference between initial and final state of system. The term $ - \int_{t_\textsf{initla}}^{t_\textsf{final}} \frac{\partial L}{\partial t}\,\dd t $ on the right hand side of the mechanical balance laws
and the term $-\frac14\frac{D^2_\textsf{L}\, c^3}{G}\int_{-\infty}^{\infty} \left(\dot{h}^2_+ + \dot{h}^2_\times\right)\mathrm{d} t$ in the balance law of general relativity represent the total energy lost by the system. In the balance law of general relativity, in the second line
of \eqref{GRBL}, there emerges an additional contribution absent in the mechanical balance laws. This is the memory effect of general relativity
\begin{equation}
\left.\frac12 \frac{D_\textsf{L}\, c^4}{G}\textsf{Re}\left[\eth^2 \left(h_+ - i\, h_\times\right)\right]\right|_{t=-\infty}^{t=+\infty} \;.
\end{equation}
Take note that the memory term essentially encapsulates the discrepancy between two distinct values: $h_{+,\times}$ in the distant future minus $h_{+,\times}$ in the distant past. In the distant past, prior to any gravitational waves being emitted, it is anticipated that 
$\left.h_{+,\times}\right|_{t=-\infty} = 0$. Likewise, in the far-off future, long after the gravitational wave has dissipated, it is foreseeable that $\left.h_{+,\times}\right|_{t=+\infty} = 0$. Thus, one naively expects $\left.h_{+,\times}\right|_{t=-\infty}^{t=+\infty} = 0$.
Nonetheless, as per the predictions of general relativity, a noteworthy revelation arises: the polarizations in the far-reaching past and the distant future may exhibit discernible discrepancies. This disparity possesses a lucid interpretation when viewed through the lens of a circular arrangement of test masses:
The propagation of a gravitational wave induces oscillatory motion within a ring of test masses. The intrinsic separation between these test masses is governed by a function dependent on $h_{+,\times}$. Subsequent to the passage of the wave, an anticipated outcome is the restoration of the original ring configuration, signifying $h_{+,\times} = 0$. However, as postulated by general relativity, it is generally the case that $h_{+,\times} \neq 0$ after the passage of the wave. Consequently, this implies a lasting alteration in the intrinsic separation between the test masses.
In turn, this means that the shape is permanently changed.
The gravitational wave memory effect, as foreseen, posits that gravitational waves culminate in an enduring displacement of test masses.

\begin{figure}[hbt!]
	\centering
	\includegraphics[width=0.8\columnwidth]{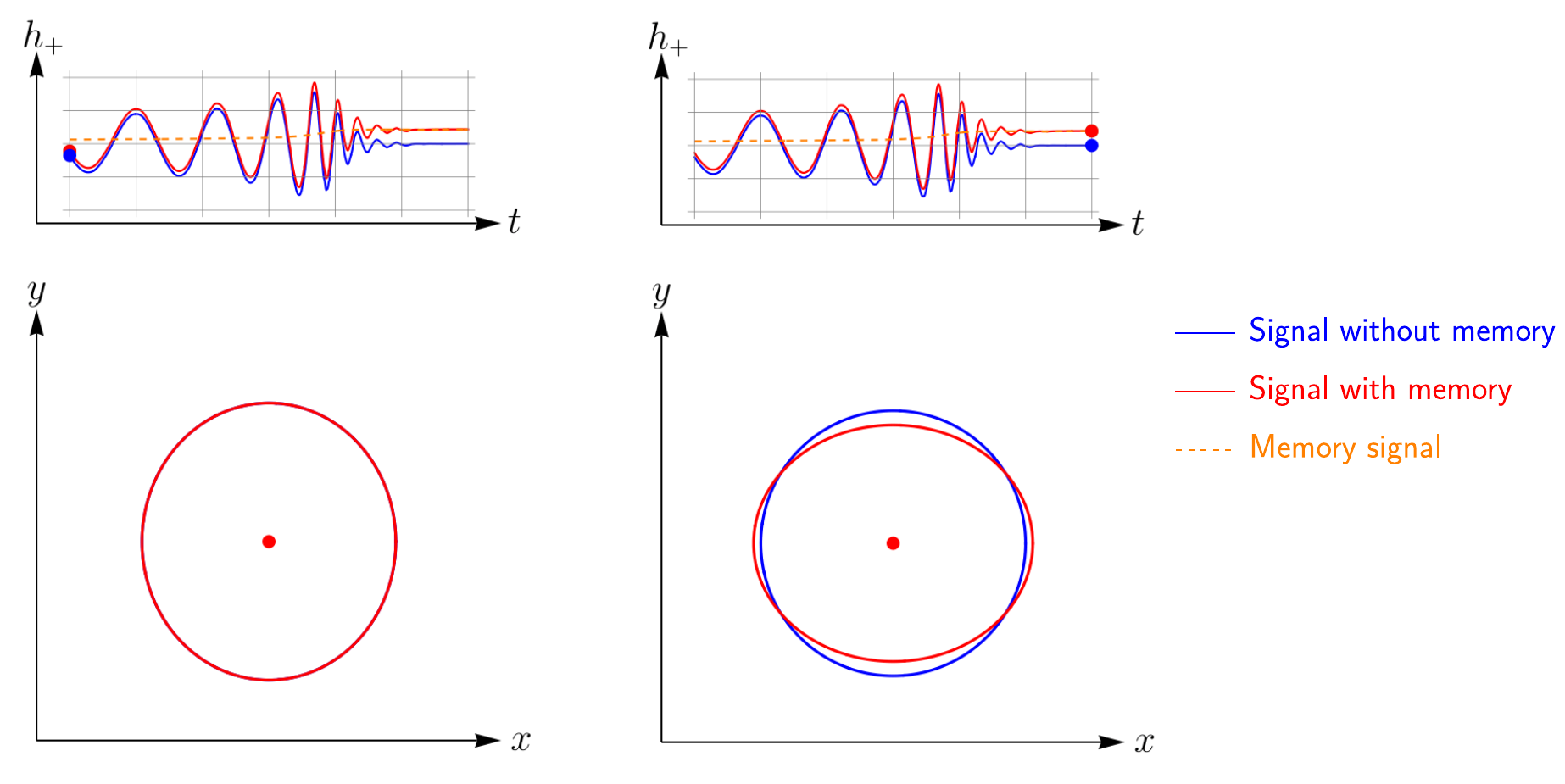}
	\caption{\protect The gravitational memory effect encapsulates the concept of permanent spacetime alterations resulting from the passage of gravitational waves. It signifies the lasting change in the separation distance between test masses after the waves have passed through, highlighting a distinctive feature of gravitational wave interactions. Left: the distribution of test masses before the passage of gravitational waves. Right: the permanent displacement after the gravitational waves have passed through. \hspace*{\fill}}
	\label{MemoryEffect}
\end{figure}

\subsection{Proof of Concept}\label{sec:ProofOfConcept}
To illustrate how the balance laws work in practice, we make use of the simple analytical waveform model devised in~\cite{Buskirk:2018ebn} for educational purposes. This shall serve as a proof of concept. This simple model assumes that there are two objects with masses $m_1$ and $m_2$, zero initial spin, and which orbit each other with a time-dependent separation $r(t)$ and a time-dependent orbital frequency $\omega(t)$ on an orbit with zero eccentricity. Furthermore, it is assumed that the orbital plane and the line-of-sight of the observer are optimally aligned. Under these simplifying assumptions, one can express the two GW polarizations during the inspiral phase as~\cite{Huerta:2016rwp}
\begin{align}
	h^{\rm ins}_+ &= -2\frac{M\, \eta}{D_{\rm L}}\left[\left(-\dot{r}^2 + r^2\,\dot{\Phi}^2 + \frac{M}{r}\right)\cos(2\Phi) + 2 r\,\dot{r}\,\dot{\Phi}\,\sin(2\Phi)\right]\notag\\
	h^{\rm ins}_\times &= -2\frac{M\, \eta}{D_{\rm L}}\left[\left(-\dot{r}^2 + r^2\,\dot{\Phi}^2 + \frac{M}{r}\right)\sin(2\Phi) - 2 r\,\dot{r}\,\dot{\Phi}\,\cos(2\Phi)\right]\,,
\end{align}
where $\Phi(t) \coloneqq \int \omega(t)\,\dd t$ is the phase of the GW, $M=m_1 + m_2$ the total mass of the binary system, $\eta = \frac{m_1\, m_2}{M^2}$ denotes the symmetric mass ratio, and $D_{\rm L}$ is the luminosity distance from the binary to the observer. The separation $r$ is approximated to 3PN order, while the phase $\Phi$ is approximated even up to 6PN order.

For the merger-ringdown phase, the authors of~\cite{Buskirk:2018ebn} use a numerical fit to NR simulations provided by~\cite{Huerta:2016rwp}. Using complex variables, the two GW polarizations can be compactly written as
\begin{align}
	h^{\rm merg}(t) = A(t)\, e^{-i\, \Phi_{\rm merg}(t)}\,,
\end{align}
where the amplitude is given by
\begin{align}
	A(t) = \frac{1}{\omega(t)} \left[\frac{|\dot{\hat{f}}|}{1 + \alpha\left(\hat{f}^2 - \hat{f}^4\right)}\right]^{\frac12}\,.
\end{align}
The function $\hat{f}$ is defined as
\begin{align}
	\hat{f} = \frac{c}{2}\left(1+\frac{1}{\kappa}\right)^{1+\kappa}\left[1-\left(1+\frac{1}{\kappa} e^{-\frac{2 t}{b}}\right)^{-\kappa}\right]\,,
\end{align}
where the parameters $b$, $c$, $\kappa$ were determined by a numerical fit to be equal to~\cite{Huerta:2016rwp}
\begin{align}
	b &= \frac{16014}{979} - \frac{29132}{1343}\eta^2\notag\\
	c &= \frac{206}{903} + \frac{180}{1141}\sqrt{\eta} + \frac{424}{1205}\frac{\eta^2}{\log(\eta)} \notag\\
	\kappa &= \frac{713}{1056} - \frac{23}{193}\eta\,.
\end{align}
The orbital frequency can also be expressed with the help of $\hat{f}$ and in terms of the spin $\hat{s}_{\rm final}$ of the remnant:
\begin{align}
	\omega(t) = \left(1-0.63\left[1-\hat{s}_{\rm final}\right]^{0.3}\right)\left(1-\hat{f}\right)\,,
\end{align}
where the final spin reads
\begin{align}
	\hat{s}_{\rm final} = 2\sqrt{3}\,\eta -\frac{390}{79}\eta^2 + \frac{2379}{287}\eta^3 - \frac{4621}{276}\eta^4\,.
\end{align}
In~\cite{Buskirk:2018ebn} it is explained in detail how the inspiral and merger-ringdown waveforms are matched and merged into a single, analytical waveform model. Using the \texttt{Mathematica} notebook provided by~\cite{Buskirk:2018ebn}, we are able to produce the following plots.
\begin{figure}[htb!]
	\centering
	\includegraphics[width=1\linewidth]{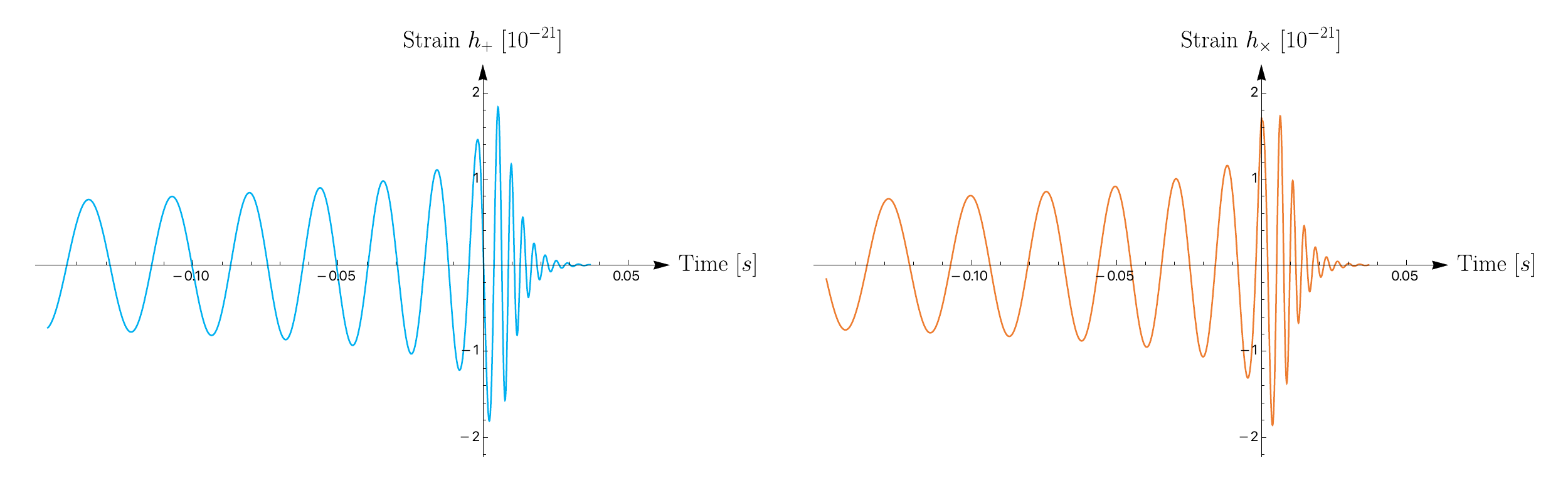}
	\caption{\protect The two waveforms generated with the help of the \texttt{Mathematica} notebook provided by~\cite{Buskirk:2018ebn} for the parameter values $m_1 = 36 M_{\odot}$, $m_2 = 29 M_{\odot}$, and $D_{\rm L} = 410 \text{Mpc}$.  \hspace*{\fill}}
	\label{fig:Waveforms}
\end{figure}
In the plots we used the data of GW150914, the first ever directly detected BH-BH merger. Specifically, we set $m_1 = 36 M_{\odot}$, $m_2 = 29 M_{\odot}$, and $D_{\rm L} = 410\, \text{Mpc}$. The mass of the remnant, as reported in~\cite{LIGOScientific:2016aoc}, is $M_{\rm remnant} = 62 M_{\odot}$. Assuming zero kick velocity, this means that an energy-equivalent of $M-M_{\rm remnant} = 3$ solar masses were radiated away during the merger.

\begin{figure}[htb!]
	\centering
	\includegraphics[width=0.75\linewidth]{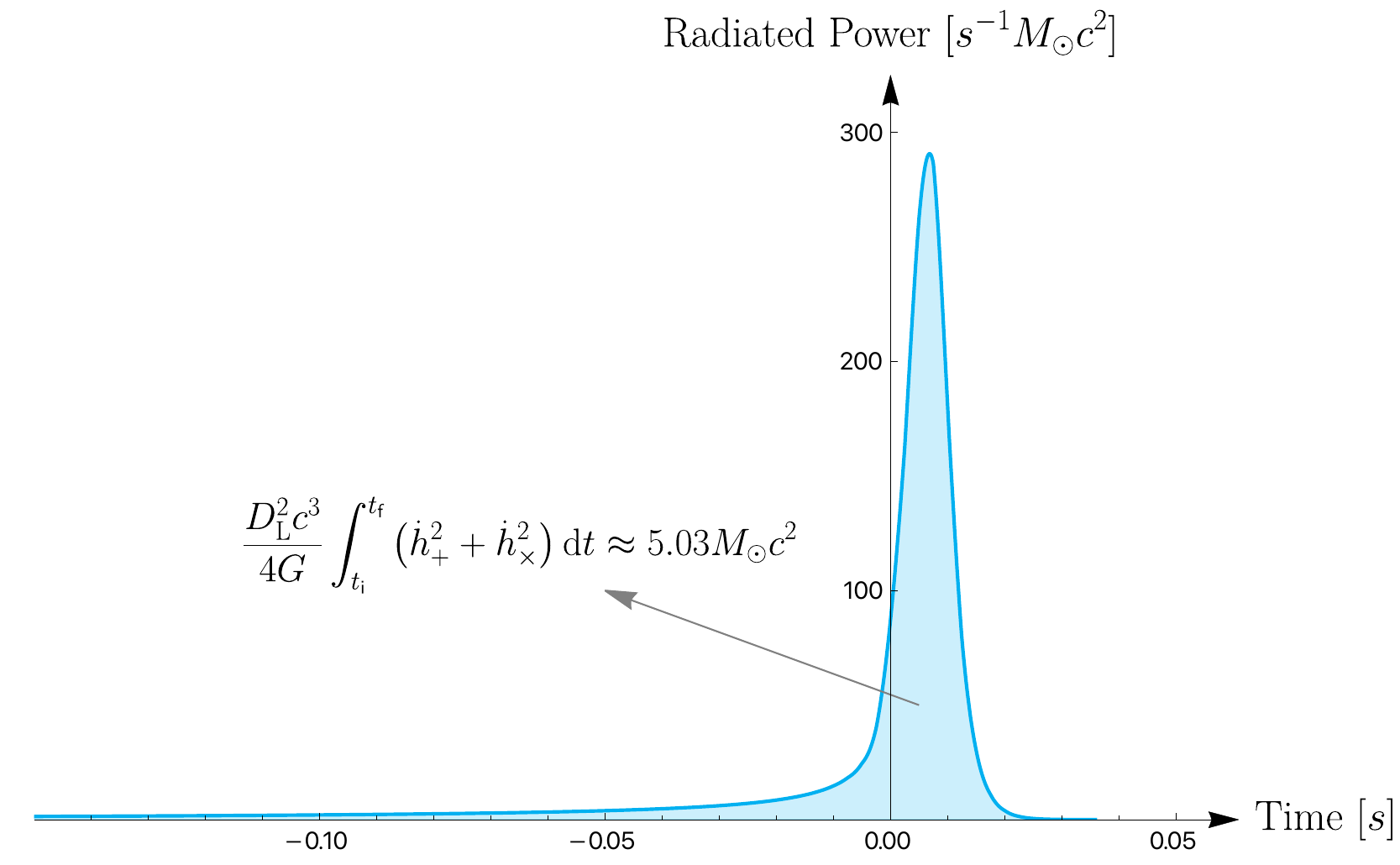}
	\caption{\protect The power radiated by GW150914 according to the model of~\cite{Buskirk:2018ebn}. The emission is strongly peaked around the merger at $t\approx 0$. Due to the strong suppression of contributions from times farther away from the merger, it is safe to truncate the energy integral. The area under the power curve corresponds to the total energy emitted during the merger and is approximately $5.03 M_{\odot} c^2$.  \hspace*{\fill}}
	\label{fig:RadiatedPower}
\end{figure}

We now also have everything we need to apply the balance laws and check the accuracy of the simple waveform model. Under the assumptions we made, the balance laws reduce to
\begin{align}
	c^2\left(M_{\rm remnant} - M\right) = -\frac{D^2_{\rm L}\, c^3}{4G}\int_{-\infty}^{+\infty}\left(\dot{h}^2_+ + \dot{h}^2_\times\right)\dd t\,,
\end{align}
where the memory term drops out and $\gamma(v_{\rm kick}) = 1$. Given the analytical waveform model, we can plot the energy-density of the GW (cf. Figure~\ref{fig:RadiatedPower}).
Evidently, the area under this curve gives us the total energy emitted during the merger. This plot also reveals that we can safely truncate the integral and evaluate it on a finite interval, rather than from $-\infty$ to $+\infty$. We find
\begin{align}
	\frac{D^2_{\rm L}\, c^3}{4G}\int_{-\infty}^{+\infty}\left(\dot{h}^2_+ + \dot{h}^2_\times\right)\dd t \approx 5.03 M_{\odot} c^2\,,
\end{align}
while the left hand side of the balance law was $c^2\left(M - M_{\rm remnant}\right) = 3 M_{\odot}c^2$. Even when taking into account the error bars on the data of~\cite{LIGOScientific:2016aoc} it is clear that the simple waveform model of~\cite{Buskirk:2018ebn} overestimates the radiated energy. It should be kept in mind, however, that the model of~\cite{Buskirk:2018ebn} was devised for educational purposes, not for accurate science. Now it also serves as a nice tool for illustrating the balance laws. This example vividly demonstrates the potency of balance laws for validating waveform models. Thus, we conclude our proof of concept.

%--------------------------------------------------------------------
%	Conclusion
%--------------------------------------------------------------------
\section{Conclusion}\label{sec:Conclusion}
Harnessing the power of balance laws, we embark on a journey to scrutinize the fidelity of gravitational waveform models within the framework of general relativity. These balance laws, elegant manifestations of the conservation principles governing physical systems, provide a rigorous yardstick to gauge the accuracy and coherence of these models. By subjecting the predicted waveforms to the scrutiny of these laws, we unveil a robust methodology for probing the compatibility of these models with the intricate fabric of spacetime dynamics. This union of balance laws and gravitational waveform models holds the potential to uncover deeper insights into the fundamental nature of gravity itself and we hold the conviction that this will truly revolutionize waveform modeling.

Evidently, at the heart of unraveling gravitational wave observations and their theoretical forecasts lies the profound importance of GW waveforms. However, the intricate endeavor of directly deducing analytical waveforms from GR persists as a formidable puzzle. Instead, these waveforms surface from a intricate amalgam of post-Newtonian theory, the effective-one-body methodology, numerical relativity, and interpolation tactics. Yet, each of these avenues introduces its own unique systematic errors, intricately entangling the synthesis process and potentially ushering in inaccuracies. As cutting-edge GW detectors gear up for activation and the rapid march of GW astronomy continues, the undeniable repercussions of these errors on research become patent. Safeguarding waveform precision emerges as a paramount directive within this evolving terrain, particularly when GW observations are harnessed to meticulously probe GR's validity in the nonlinear realm. The peril of veiling deviations from GR, or conversely, fallaciously identifying them due to waveform intricacies, accentuates the compelling necessity for an autonomous mechanism to authenticate waveform consistency.

Each approximation technique inevitably introduces errors that can affect the accuracy of the waveform. Considering the vital role of waveform models in detecting gravitational wave events and extracting valuable information from the signals, a crucial question arises: Which waveform model best approximates the predictions of full, non-linear General Relativity regarding the physics of compact binary coalescence? This quest can be addressed through the application of balance laws. To elucidate the potential of balance laws, we initiated our exploration with a detailed mechanical analogy. We utilized a dissipative mechanical system, exemplified by a ball's descent down a frictional half-pipe, characterized by a non-linear equation. This underscores the role of mechanical balance laws as a benchmark for assessing the precision of approximate solutions in encapsulating the complete physical picture. Employing a small-angle approximation, we derived an approximate analytical solution and assessed its accuracy in representing the fundamental physics through a comparison with the unapproximated mechanical balance laws. As anticipated, the fidelity of the balance laws holds true primarily within the realm of small angles, where the linear approximation effectively serves as a reliable solution.

Continuing, we delved into the distinctive challenges inherent in deriving balance laws in electromagnetism and general relativity, deviating from the direct mechanical formulation. This endeavor necessitated a robust foundation rooted in mathematical concepts of radiation. We initiated with an electromagnetism analogy, introducing pivotal concepts such as conformal completion, the Newman–Penrose formalism, Carter–Penrose diagrams, and the profound Peeling Theorem. Subsequently, with the complete set of balance laws in general relativity at our disposal, we employed them as a litmus test to assess the precision and validity of a specific approximate waveform model~\cite{Buskirk:2018ebn}, substantiating its feasibility as a proof of concept.

\vskip6pt

\ack{LH is supported by funding from the European Research Council (ERC) under the European Unions Horizon 2020 research and innovation programme grant agreement No 801781. LH further acknowledges support from the
Deutsche Forschungsgemeinschaft (DFG, German Research Foundation) under Germany's Excellence Strategy EXC 2181/1 - 390900948 (the Heidelberg STRUCTURES Excellence Cluster).}

%%%%%%%%%% Insert bibliography here %%%%%%%%%%%%%%

\end{document}